\documentclass[pre,aps,onecolumn,superscriptaddress,notitlepage,nofootinbib]{revtex4-1}
\usepackage{amssymb,epsfig,amsmath,bm,subfigure,graphicx,textcomp,url,float,color,cases}
\usepackage[colorlinks=true, urlcolor=blue, anchorcolor=blue, citecolor=blue,filecolor=blue,linkcolor=blue,menucolor=blue]{hyperref}

\usepackage[titletoc,title]{appendix}

\begin{document}

\title{Short-time large deviations of the spatially averaged height\\of a KPZ interface on a ring}

\author{Timo Schorlepp}
\email{timo.schorlepp@rub.de}
\affiliation{Institute for Theoretical Physics I,
Ruhr University Bochum, 44801 Bochum, Germany}

\author{Pavel Sasorov}
\email{pavel.sasorov@gmail.com}
\affiliation{ELI Beamlines Facility,
ERIC, 25241 Doln\'{\i} B\u{r}e\v{z}any, Czech Republic}

\author{Baruch Meerson}
\email{meerson@mail.huji.ac.il}
\affiliation{Racah Institute of Physics, Hebrew
  University of Jerusalem, Jerusalem 91904, Israel}

\date{\today}

\begin{abstract}
Using the optimal fluctuation method, we evaluate the short-time probability
distribution $P (\bar{H}, L, t=T)$ of the spatially averaged height $\bar{H} = (1/L) \int_0^L h (x, t=T)\,dx$
of a one-dimensional interface $h (x, t)$ governed by the Kardar--Parisi--Zhang equation
$$
\partial_th=\nu \partial_x^2h+\frac{\lambda}{2}
\left(\partial_xh\right)^2+\sqrt{D}\xi(x,t)
$$
on a ring of length $L$. The process starts from a flat interface, $h(x,t=0)=0$.
Both at $\lambda \bar{H}<0$, and at sufficiently small positive $\lambda \bar{H}$ the optimal
(that is, the least-action) path $h(x,t)$ of the interface, conditioned on $\bar{H}$, is uniform
in space, and the distribution $P (\bar{H}, L, T)$ is Gaussian. However, at sufficiently
large $\lambda \bar{H}>0$ the spatially uniform solution becomes sub-optimal and gives way
to non-uniform optimal paths. We study them, and the resulting non-Gaussian distribution $P (\bar{H}, L, T)$,
analytically and numerically. The loss of optimality of the uniform solution occurs via a dynamical
phase transition of either first, or second order, depending on the rescaled system size
$\ell = L/\sqrt{\nu T}$, at a critical value $\bar{H}=\bar{H}_{\text{c}}(\ell)$. At large but
finite $\ell$ the transition is of first order. Remarkably, it becomes an ``accidental" second-order
transition in the limit of $\ell \to \infty$, where a large-deviation
behavior $-\ln P (\bar{H}, L, T) \simeq (L/T) f(\bar{H})$
(in the units $\lambda=\nu=D=1$) is observed. At small $\ell$ the transition is of second order,
while at $\ell =O(1)$ transitions of both types occur.

\end{abstract}

\maketitle
\tableofcontents


\section{Introduction}
Atypically large fluctuations in macroscopic systems out of
equilibrium continue to attract great interest from statistical physicists.
Although a universal description of such fluctuations is unavailable, there has been
much progress in studies of particular systems. One of the main theoretical tools
in this area is known under different names in different areas of physics:
the optimal fluctuation method
(OFM), the instanton method, the weak-noise theory, the
macroscopic fluctuation theory, \textit{etc}. This method relies
on a saddle-point evaluation of the pertinent path integral
of the stochastic process, conditioned on the
large deviation. The method is based on a model-specific
small parameter (often called ``weak noise"), and it brings about a
conditional variational problem. The solution of this problem -- a
deterministic, and in general time-dependent, field -- describes the ``optimal path" of the system:
the most probable system's history which dominates the contribution of different
paths to the statistics in question.

Among multiple applications of the OFM, we focus on one set of problems which has attracted attention in the last
two decades~\cite{KK2007,KK2008,KK2009,MKV,KMSparabola,LDMRS,Janas2016,KLD2017,MeersonSchmidt2017,SMS2018,SKM2018,
SmithMeerson2018,Hartmann2018,MV2018,Asida2019,SMV2019,HMS2019,KLD2021,HMS2021,KLD2022,Lamarre,SGG}: short-time
large deviations of a stochastically growing interface as described by the one-dimensional Kardar--Parisi--Zhang (KPZ) equation~\cite{KPZ}
\begin{equation}\label{BH010}
\partial_th=\nu \partial_x^2h+\frac{\lambda}{2}
\left(\partial_xh\right)^2+\sqrt{D}\xi(x,t)\, ,
\end{equation}
where $\xi(x,t)$ is a white noise with
\begin{equation}\label{BH020}
\langle\xi(x,t)\rangle=0\,,\quad \langle\xi(x,t)\xi(x^\prime,
t^\prime)\rangle=\delta(x-x^\prime)\delta(t-t^\prime)\,.
\end{equation}
Here we employ the OFM to study a KPZ interface on a ring of length $L$, \textit{i.e.}\ with periodic boundary
conditions at $x=0$ and $x=L$. The interface is initially flat,
\begin{equation}\label{BH040stoch}
h(x,t=0)=0\,,
\end{equation}
and we are interested in evaluating
the probability density function (PDF) $P\left(\bar{H}, L, T\right)$
of the spatially averaged surface height
\begin{equation}\label{BH030}
\bar{H} = \frac{1}{L}\int_0^L h(x,T)\, dx
\end{equation}
at a final time $t=T >0$, which is much shorter than the characteristic nonlinear
time of Eq.~(\ref{BH010}), $\tau_{\text{NL}}= \nu^5/D^2 \lambda^4$.
The short-time limit allows one to employ the OFM in a controlled
manner~\cite{KK2007,KK2008,KK2009,MKV,KMSparabola,Janas2016,MeersonSchmidt2017,SMS2018,SKM2018,
SmithMeerson2018,MV2018,Asida2019,SMV2019,HMS2019,KLD2021,HMS2021,KLD2022,SGG}, as we will
reiterate shortly. The problem, defined by Eqs.~(\ref{BH010})-(\ref{BH030}), continues the
line of studies of Refs.~\cite{SMS2018,SGG} of finite system-size effects (which turn out to be quite dramatic)
in large deviations of height of the KPZ interface.

Upon rescaling $t \to tT$,
$x \to (\nu T)^{1/2} x$, $h \to \nu h / \lambda$ and $\xi \to
\left(\nu T^3\right)^{-1/4} \xi$, Eq.~(\ref{BH010}) becomes
\begin{equation}\label{BH050stoch}
\partial_th= \partial_x^2h+\frac{1}{2}\left(\partial_xh\right)^2
+\sqrt{\varepsilon}\xi(x,t)\,,
\end{equation}
with rescaled noise strength $\varepsilon = D \lambda^2 T^{1/2}
/ \nu^{5/2}$ on a ring of rescaled length $\ell = L / \sqrt{\nu T}$.
The PDF of the rescaled average height $\bar{H}$ at final time $t = 1$
can then be written as a path integral
\begin{equation}
\label{pathint}
P(\bar{H},\ell,\varepsilon) = \int_{h(\cdot, 0) = 0} Dh \; \delta \left(
\frac{1}{\ell} \int_0^\ell h(x,1) \, dx - \bar{H} \right)
J[h] \exp \left\{-\frac{1}{\varepsilon} S[h] \right\}
\end{equation}
with action functional
\begin{align}\label{eq:action-func}
S[h] = \int_0^1 dt \int_0^\ell dx \, {\cal L}
\left(h, \partial_t h \right) = \frac{1}{2} \int_0^1 dt
\int_0^\ell dx \left[\partial_th - \partial_x^2h-\frac{1}{2}
\left(\partial_xh\right)^2 \right]^2\,,
\end{align}
where $\mathcal{L}(h,\partial_t h)$ is the Lagrangian.
The OFM assumes a weak-noise limit $\varepsilon\to 0$, when the path integral~(\ref{pathint}) can be evaluated
by the saddle-point method, while the Jacobian $J[h]$ does not contribute in the leading-order.
In this limit, the PDF $P(\bar{H},\ell,\varepsilon)$ is dominated by
the optimal path of the system, that is by the most likely history $h(x,t)$ conditional on a given average height at $t=1$:
\begin{align}\label{eq:pdf-min}
-\ln P(\bar{H}, \ell, \varepsilon) \overset{\varepsilon \to 0}{\simeq}
\varepsilon^{-1} \min_{\substack{h(\cdot, 0)= 0\,,\\\int_0^\ell
h(x,1)dx = \ell\bar{H}}} S[h] = \varepsilon^{-1} S(\bar{H}, \ell)\,.
\end{align}
Hence, the PDF can be determined, up to pre-exponential factors,  from the
solution of this constrained minimization problem. Here
we will solve this minimization problem numerically, for different $\bar{H}$
and $\ell$, and analytically in the asymptotic limits of large and small
$\ell$\footnote{\label{footnote:whenever}Note that whenever there exists a spatially
non-uniform optimal path, there are actually infinitely many possible
paths due to the translational symmetry of the problem with respect to $x$. Accounting for
this submanifold of degenerate solutions and for the associated zero
mode is, however, only relevant for pre-exponential factors~\cite{SGG} which
we do not address here.}.

It will be convenient to present our results by setting
$\nu=\lambda=D=1$\footnote{\label{footnote:lambdapositive}In most of the paper we assume, without
loss of generality, that $\lambda>0$. Indeed, changing $\lambda$ to $-\lambda$ is equivalent to changing $h$ to $-h$.}.
Then the weak-noise scaling~(\ref{eq:pdf-min}) reads
\begin{align}\label{eq:pdf-min1}
-\ln P(\bar{H}, \ell, \varepsilon\to 0) \simeq
 T^{-1/2} S(\bar{H}, \ell)\,.
\end{align}
Note that the limit $\varepsilon \to 0$ at fixed $\ell$ corresponds to
the short-time limit $T \to 0$ and small-length limit $L \to 0$
with $L / \sqrt{T} = \text{const}$.
When instead $T$ goes to zero at $L=\text{const}$, one has
both $\varepsilon \to 0$ and $\ell
\to \infty$. The latter limit turns out to be most interesting, and it is analyzed here
in detail. It is natural to expect that for
any $\bar{H}$, when $\ell\to\infty$, the action $S(\bar{H}, \ell)$ should exhibit
a large-deviation form
\begin{equation}\label{BH120}
S(\bar{H},\ell) \overset{\ell \to \infty}{\simeq} \ell f(\bar{H})\,,
\end{equation}
leading to
\begin{align}\label{eq:pdf-min2}
-\ln P(\bar{H}, L, T\to 0) \simeq
(L/T) \,f(\bar{H})\,,
\end{align}
and this is what we indeed observe here. Less expectedly, we also find that the rate
function $f(\bar{H})$ exhibits,  at a critical value $\bar{H}=\bar{H}_{\text{c}}(\ell)$,
a dynamical phase transition (DPT) which is \emph{accidentally} second-order.
By that we mean that
the rate function at the critical point becomes continuously differentiable
\emph{only} in the limit of $\ell\to \infty$. At arbitrary large but finite $\ell$ the
large-deviation form~(\ref{BH120}) breaks down. We show, however, that the action $S(\bar{H},\ell)$ still exhibits
a DPT at a critical point $\bar{H}=\bar{H}_{\text{c}}$, but this DPT is of (weakly) first order and the optimal
path at the critical point changes discontinuously via a subcritical bifurcation.

\begin{figure}
\centering
\includegraphics[width=.58\textwidth]{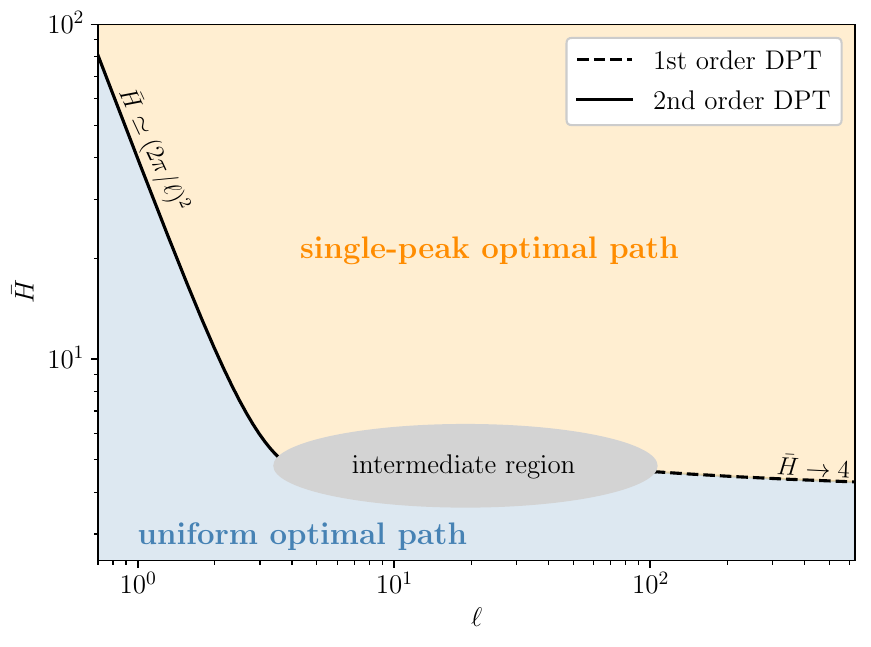}
\caption{An overview of our main results for the
large-deviation problem~(\ref{eq:pdf-min}) on the $(\bar{H},\ell)$ phase diagram.  
In the light blue region, which extends to all $\bar{H} < 0$, the global minimum of the action
functional~(\ref{eq:action-func}) is attained at the spatially uniform
solution~(\ref{BH210}). In the light orange region, the least-action solution $h(x,t)$
is instead a spatially non-uniform path with a single maximum. For small domain
sizes $\ell \to 0$ it is given by Eq.~(\ref{eq:h-ellip}), and for large domain
sizes it is described
in Sec.~\ref{sec:hLg}. For small~$\ell$,
there is a second order DPT from the uniform to the peaked solution, as discussed
in Sec.~\ref{un} and Sec.~\ref{sec:small-ell}. For large~$\ell$, a (weakly) first
order transition takes place instead, at~$\bar{H}$ approximately given by Eq.~(\ref{D440}).
In the intermediate region 
transitions of both first and second order occur. Here
some numerical results are available, see Fig.~\ref{fig:branch2-transition-diff-l}, but a complete understanding
is lacking. 
}
\label{fig:phase-diag}
\end{figure}

For small $\ell$ a truly second-order DPT is observed as predicted earlier~\cite{SMS2018,SGG}.
At intermediate values of $\ell = O(1)$ DPTs of both types occur. In the latter regime analytical
results are unavailable as of yet, and we present some numerical results. All the DPTs that we
found in this system occur because of a loss of optimality of a path that is uniform in space.
The loss of optimality takes the form either of a subcritical bifurcation (for the first-order DPTs),
or a supercritical bifurcation (for the true second-order DPTs). We summarize our main results on the $(\bar{H},\ell)$ phase diagram of the system in Fig.~\ref{fig:phase-diag}.

The remainder of this paper is structured as follows. In Sec.~\ref{OFMeqn} we formulate
the OFM equations and boundary conditions, present a simple uniform solution of these equations,
previously studied in Refs.~\cite{SMS2018,SGG}, and
argue that it describes the optimal path of the system at all $\lambda H<0$.  Supercritical
bifurcations of the uniform solution have been recently studied in Ref.~\cite{SGG}. Still,
for convenience of further discussion, we briefly rederive them in Sec.~\ref{un}.
Section~\ref{sec:num} includes our results of numerical minimization of the action
functional~(\ref{eq:action-func}) in different regions of the $(\bar{H},\ell)$ phase diagram.
These numerical results provided valuable insights into the nature of optimal paths of the
interface which led us to develop asymptotic analytical solutions of the OFM problem for
large $\ell$ that we present in Sec.~\ref{sec:hL}.  The asymptotic solution for small $\ell$
is briefly discussed in Sec.~\ref{sec:small-ell}. We summarize and discuss our main results
in Sec.~\ref{summary}. A description of numerical algorithms that we use here is relegated to the Appendix.


\section{OFM equations and uniform solution} \label{OFMeqn}
At a technical level, the main objective of this work is to determine the minimum action $S(\bar{H}, \ell)$
as a function of the rescaled average height $\bar{H}$ and rescaled
system size $\ell$. In this section, we present the necessary
conditions for minimizers of the action functional~(\ref{eq:action-func}) -- the OFM equations and the boundary conditions.
We argue then that a simple  spatially uniform solution
of the ensuing OFM problem is always optimal for $\bar{H} < 0$.

The first-order necessary conditions for a minimizer of the action
functional~(\ref{eq:action-func}) can be represented as a pair of Hamilton's equations
for the optimal history of the interface $h(x,t)$ and the
conjugate momentum density $p = \partial {\cal L} / \partial(\partial_t h)$. These equations
were derived in many papers~\cite{KK2007,KK2008,KK2009,MKV,KMSparabola,Janas2016,MeersonSchmidt2017,SMS2018,SKM2018,
SmithMeerson2018,MV2018,Asida2019,SMV2019,HMS2019,KLD2021,HMS2021,KLD2022,SGG}, and they take the form
\begin{eqnarray}
 \partial_th&=&\partial_x^2h+\frac{1}{2}\left(\partial_xh\right)^2+p
 \,,\label{BH050}\\
\partial_tp&=&-\partial_x^2p+\partial_x\left(p\partial_xh\right)
\,. \label{BH060}
\end{eqnarray}
The ``momentum density" $p(x,t)$ describes the (rescaled) optimal realization of
the external noise $\xi(x,t)$ that drives the interface conditional on a specified $\bar{H}$.
In the present case Eq.~(\ref{BH050}) and~(\ref{BH060}) should be complemented by the periodic  boundary conditions
at $x=0$ and $x = \ell$, by the initial condition
\begin{equation}\label{BH040}
h(x,0)=0\,,
\end{equation}
and by the final-time condition
\begin{equation}\label{BH070}
p(x,1)=\Lambda=\mbox{const}\,,
\end{equation}
which follows from the demand that a boundary term at $t=1$, originating from an
integration by parts, should vanish for any $h(x,1)$.
The parameter $\Lambda$ is a Lagrange multiplier which needs
to be chosen so as to impose the rescaled final-time condition
\begin{equation}\label{eq:final-constr}
\frac{1}{\ell} \int_0^\ell h(x,1) dx = \bar{H}\,.
\end{equation}
Once the optimal path is determined, the action $S(\bar{H},\ell)$
can be determined from the equation
\begin{equation}\label{action01}
S = \frac{1}{2}\int\limits_0^1 dt\int\limits_0^\ell dx\, p^2(x,t)\,,
\end{equation}
which follows from Eqs.~(\ref{eq:action-func}) and~(\ref{BH050}).

By differentiating the action $S(\bar{H}, \ell) = S[h(x,t;\bar{H},\ell)]$ of
the optimal profile $h = h(x,t;\bar{H},\ell)$ with respect to $\bar{H}$ using
the chain rule, one can show that $\Lambda$ is related to the action via
\begin{equation}\label{BH092}
\Lambda=\frac{1}{\ell}\,\frac{\partial S(\bar{H}, \ell)}{\partial \bar{H}}
\mbox{~~~~(or~~~~} dS=\ell\Lambda d\bar{H})\,.
\end{equation}
If the action $S(\bar{H}, \ell)$ is a strictly convex function of $\bar{H}$,
there is a bijective relation between $\Lambda$ and $\bar{H}$, and it
suffices, for the purpose of calculating the action, to only
determine $\bar{H}(\Lambda)$ and use Eq.~(\ref{BH092}). This shortcut is very convenient and
holds for many large-deviation calculations~\cite{shortcut}.

There is an obvious exact solution of the OFM equations and the boundary conditions:
\begin{equation}\label{BH210}
h(x,t)=\bar{H} t\,,\qquad p(x,t)=\Lambda\,,\qquad \Lambda = \bar{H}\,,
\qquad S=\frac{\ell}{2} \bar{H}^2\,,
\end{equation}
which describes a uniformly growing flat interface.
We will often call this branch of solutions branch 1. By virtue of Eq.~(\ref{eq:pdf-min}),
whenever the uniform solution~(\ref{BH210}) is the optimal one, we have
a Gaussian PDF for $\bar{H}$ up to pre-exponential factors. Of most interest, however,
are the regions of parameters $\bar{H}$
and $\ell$, for which the uniform solution is sub-optimal. As we will see,
the loss of optimality can occur via either a supercritical, or a subcritical bifurcation.

First of all, we can argue that, for negative $\bar{H}$, the uniform
solution~(\ref{BH210}) is always optimal. Using the evident conservation law
\begin{equation}\label{conservationp}
\frac{1}{\ell} \int_0^\ell p(x,t) \,
d x = \Lambda = \text{const}
\end{equation}
of Eq.~(\ref{BH060}), we can rewrite the action~(\ref{eq:action-func}) for any solution
of the OFM equations as
\begin{equation}\label{BH080}
S = \frac{1}{2}\int\limits_0^1 dt\int\limits_0^\ell
dx\, p^2(x,t)=\ell\frac{\Lambda^2}{2}+\frac{1}{2}
\int\limits_0^1 dt\int\limits_0^\ell dx\,
[p(x,t)-\Lambda]^2\,,
\end{equation}
Also, integrating both sides of Eq.~(\ref{BH050}) with respect to $t$ from $0$ to $1$ and
with respect to $x$ over the ring, and using the periodic boundary conditions
and the conservation law~(\ref{conservationp}), we obtain
\begin{equation}\label{BH090}
\bar{H}=\frac{1}{\ell}\int\limits_0^\ell h(x,1) \, dx
=\Lambda+\frac{1}{2\ell}\int\limits_0^1 dt\int\limits_0^\ell
dx\,\left[\partial_xh(x,t)\right]^2\,.
\end{equation}
One can easily see from Eqs.~(\ref{BH080}) and~(\ref{BH090})  that, at negative $\Lambda$
(or $\bar{H}$) any inhomogeneity in the
momentum density $p$ both increases
the action $S$, and decreases the average height $|\bar{H}|$ in comparison to their
values for the uniform solution. Therefore, any nonuniform solution here is sub-optimal.

In contrast to this, for $\Lambda >0$ (or
$\bar{H}>0$), an inhomogeneity  increases both $S$,
and $\bar{H}$ in comparison to the uniform solution. A competition
between these two opposite effects may give rise to non-uniform solutions with lesser action than
the uniform one, as we will indeed see in the following.


\section{Bifurcations of the uniform solution} \label{un}
In this brief section we carry out a linear stability analysis of the
uniform solution~(\ref{BH210}). We find that, for sufficiently
large positive $\bar{H}$, the uniform solution can continuously
and supercritically bifurcate to a non-uniform solution. The first
spatial Fourier mode to become unstable as $\bar{H}$ increases depends
on the rescaled system size $\ell$ in a nontrivial way and is determined
from Eq.~(\ref{BH250}). This equation has also been obtained in Ref.~\cite{SGG}
by calculating the leading-order prefactor correction to the asymptotic
scaling in Eq.~(\ref{eq:pdf-min}) through Gaussian integration of
fluctuations around the uniform solution~(\ref{BH210}).

\begin{figure}
\centering
\includegraphics[width=.4\textwidth]{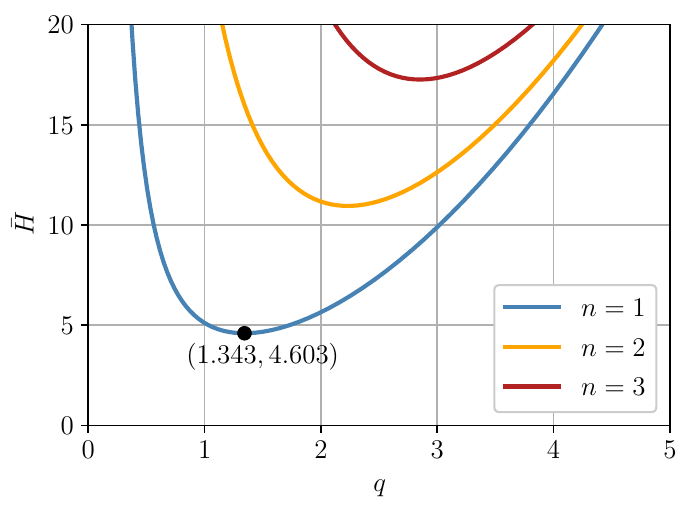}
\caption{The supercritical bifurcation
points~$\bar{H} = \bar{H}_{\text{c}}$ vs.\ the wave
number~$q$ as predicted by Eq.~(\ref{BH260}) for $n=1,2$ and $3$.  Only discrete values
of $q=2 \pi m/\ell$ with $m = 1, 2, \dots$ are allowed. The lowest curve corresponds to $n=1$.
The black dot indicates the global minimum of $\bar{H}_{\text{c}}$ vs.\ $q$, so
that $\bar{H}_{\text{c}} \geq 4.603$ is a necessary condition for a supercritical bifurcation
of the uniform solution for any~$\ell$.}
\label{fig:barhcrit}
\end{figure}

At first order of a perturbation theory around the uniform
solution~(\ref{BH210}) we have
\begin{equation}\label{BH220}
p(x,t)=\bar{H}+b(t)\cos qx\,,\qquad h(x,t)=\bar{H} t + a(t)\cos qx
\,,\qquad |a|,\, |b|\ll 1\,.
\end{equation}
Here the wave number $q$ spans the set $2\pi m/\ell$ for
$m=1,2,\dots$. Substituting the expressions~(\ref{BH220})
into Eqs.~(\ref{BH050}) and~(\ref{BH060}) and neglecting higher-order terms, we obtain
the following system
of linear ordinary differential equations:
\begin{equation}\label{BH230}
\dot{a}=-q^2a+b\,,\qquad \dot{b}=q^2b-q^2\bar{H} a\,.
\end{equation}
It has solutions proportional to $e^{i\omega t}$, where
\begin{equation}\label{BH240}
\omega=\pm q \sqrt{\bar{H}-q^2}\,.
\end{equation}
Using the boundary conditions~(\ref{BH040}) and~(\ref{BH070}), we obtain the
following relationship between $q$ and $\bar{H} = \bar{H}_\text{c}(q)$
at the bifurcation points:
\begin{equation}\label{BH250}
\tan \left(q\sqrt{\bar{H}-q^2}\right)=-\frac{\sqrt{\bar{H}-q^2}}{q}\,.
\end{equation}
Note that the trivial solution $\bar{H}=q^2$ of Eq.~(\ref{BH250}) does
not correspond to a valid non-uniform solution due to the boundary conditions
at $t=0$ and $1$. The resulting dependence $\bar{H}(q)$ can be expressed in a
parametric form
\begin{equation}\label{BH260}
\bar{H} = -\frac{2 u}{\sin 2u}\,,\qquad q=\sqrt{-u\cot u}\,,
\qquad \frac{(2n-1)\pi}{2}<u<n\pi;\quad n=1,2,3,\dots\,,
\end{equation}
where, for given $\ell$, only values of $q = 2 \pi m / \ell$
with $m = 1, 2, 3, \dots$ are allowed.
The first three branches of Eq.~(\ref{BH260}) are shown in
Fig.~\ref{fig:barhcrit}. As one can see, the first instability appears for $n = 1$,
and a necessary condition for the instability, for any $\ell$, is $\bar{H}_{\text{c}} \geq 4.603$.
When $\ell \to \infty$, the first instability of the
uniform solution will occur, at $\bar{H}_{\text{c}} \simeq 4.603$, for a very high mode
$m \simeq 1.343 \,\ell/ 2 \pi$.
For finite $\ell$, one can find the bifurcation point on the $n=1$ branch of Eq.~(\ref{BH260})
numerically.
Finally, for $\ell \to 0$, the first instability occurs for the $m = 1$ mode at
$\bar{H} \simeq (2 \pi / \ell)^2$ in
agreement with Ref.~\cite{SMS2018}.

\section{Numerical results} \label{sec:num}
Now we proceed with a numerical solution of the
minimization problem in Eq.~(\ref{eq:pdf-min}) for different $\bar{H}$ and $\ell$. The numerical methods that
we used are described in the Appendix. In addition to confirming
the supercritical bifurcations of the uniform solution that we discussed in Sec.~\ref{un},
we will uncover important subcritical bifurcations
and get insight into non-perturbative optimal paths which
will be studied analytically in Secs.~\ref{sec:hL} and~\ref{sec:small-ell}.

\begin{figure}
\includegraphics[width = .8\textwidth]{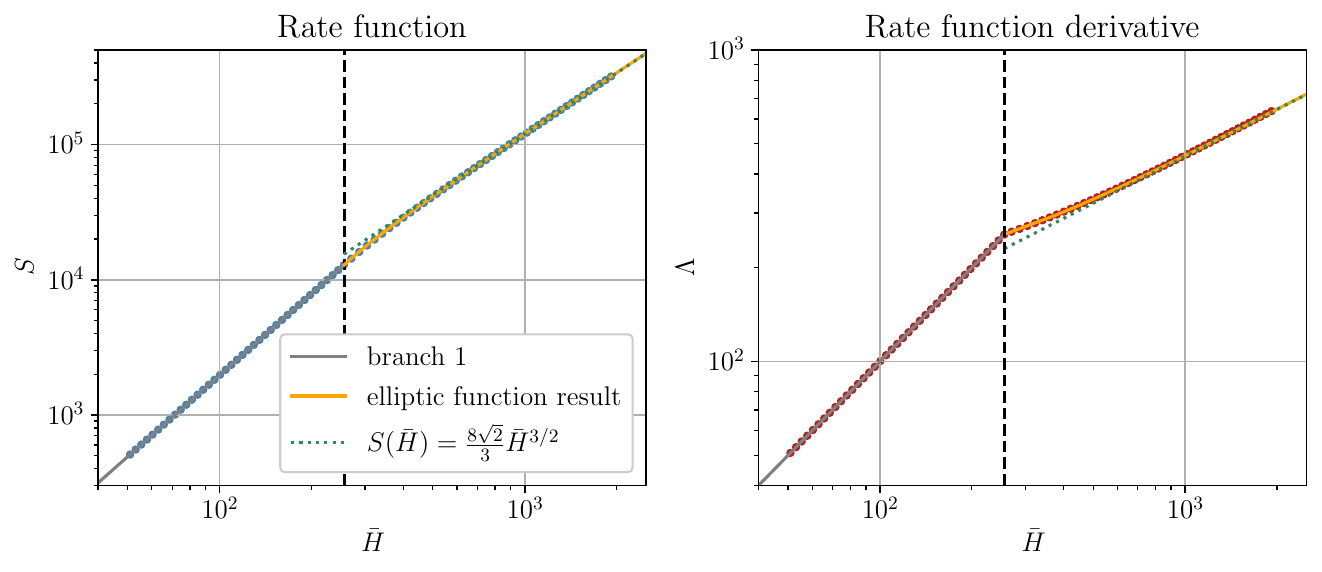}
\caption{Comparison of analytical small-$\ell$ results (lines)
and numerics (dots)
for the rate function $S(\bar{H})$ (left) and Lagrange
multiplier $\Lambda(\bar{H})$ (right) for a rescaled system size
of $\ell = \pi / 8$. The
numerical computations were performed at
resolutions $n_x = 64$ and $n_t = 8000$.
The dashed vertical line indicates where
the spatially uniform solution becomes unstable
at $\bar{H} \simeq 256.04$ in agreement with Eq.~(\ref{BH250}).
The dotted green line
corresponds to the asymptotic $S(\bar{H}) =(8\sqrt{2}/3)\,\bar{H}^{3/2}$, see Eq.~(\ref{eq:action-asymp}),
the gray line corresponds to the spatially uniform
solution~(\ref{BH210}), and the orange line corresponds to Eqs.~(\ref{eq:ellip-action}) and~(\ref{Hbarvsk}).}
\label{fig:l-0125-pi-rf-lbda}
\end{figure}

\begin{figure}
\includegraphics[width = \textwidth]{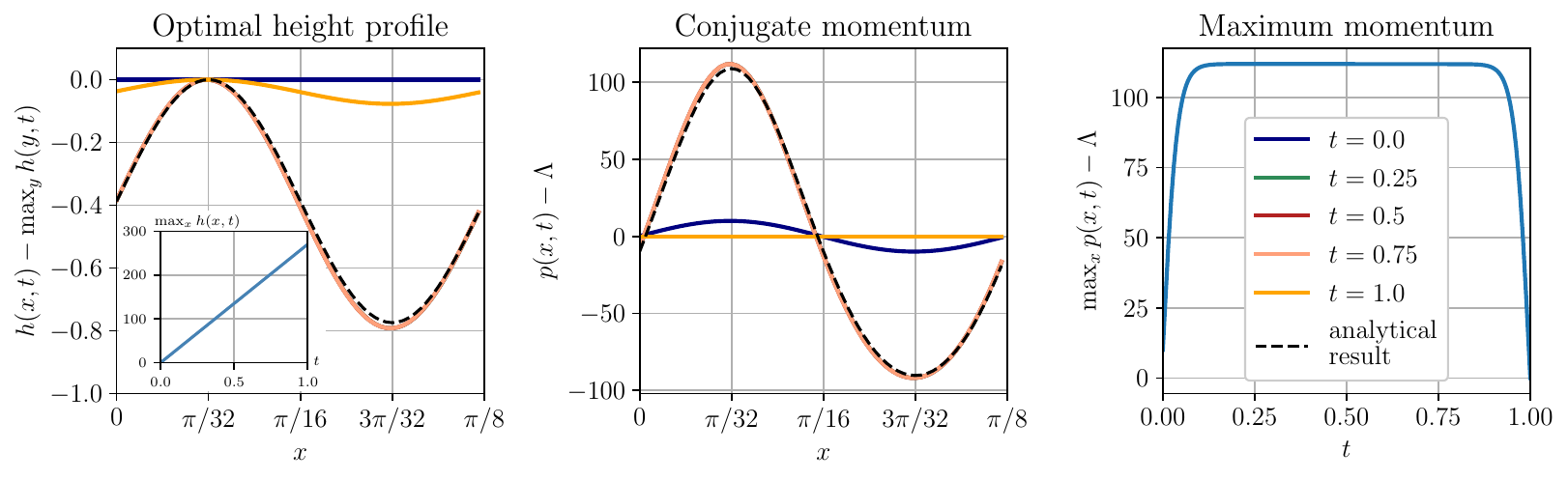}
\caption{Numerically found spatially slightly non-uniform
solution of Eqs.~(\ref{BH050})
and~(\ref{BH060}) for $\bar{H} = 270.319$ and
a moderately small rescaled system size of $\ell = \pi / 8$ with
numerical resolution $n_x = 64$ and $n_t = 8000$.
The numerically found action  $S = 14322.081$
deviates by $0.009\%$ from the predicted small-$\ell$
result $S = 14320.806$ for this $\bar{H}$ as given by Eq.~(\ref{eq:ellip-action}).
The action for the
spatially uniform solution~(\ref{BH210}) for the same $\bar{H}$ and $\ell$
would be $S = 14347.685$.
The maximum height at final time is $H = 270.357$.
Left: Optimal height profile $h(x,t)$ at different
times $t$, with the maximum at each $t$ subtracted in order
to emphasize deviations from spatial homogeneity.
The prediction~(\ref{eq:h-ellip})
for intermediate times is indicated by the black dashed line
and agrees well with the numerical solution for $t = 0.25$,
$t = 0.5$ and $t = 0.75$. The inset shows that the growth of
the maximum $\max_x h(x,t)$ in time is still linear as predicted.
Center: conjugate
momentum density $p(x,t)$ with $\Lambda = 261.057$
subtracted, compared to the analytical
result~(\ref{eq:p-ellip}) indicated
by the black dashed line. Right: Spatial
maximum $\max_x p(x,t)$ over time to visualize the long lifetime of
the stationary cnoidal solution with some small boundary layers in time at $t=0$ and $t=1$.}
\label{fig:small-l-solution}
\end{figure}

We start with the simpler case of small $\ell$.
Choosing a moderately small value $\ell = \pi / 8$ and numerically
minimizing the action~(\ref{eq:action-lbda}) for different $\Lambda$, we
obtain
the rate function $S(\bar{H}, \ell)$ and Lagrange
multiplier $\Lambda(\bar{H})$ shown in Fig.~\ref{fig:l-0125-pi-rf-lbda}.
The spatially uniform solution~(\ref{BH210}), corresponding
to branch 1 of the action, is seen to become unstable
close to $\bar{H} \simeq (2 \pi  / \ell)^2$ as stated in Sec.~\ref{un},
and there is a
continuous (second-order) DPT to a spatially
nonuniform solution. Indeed, the $(m = 1)$-spatial Fourier mode of the
profile becomes unstable at this point.  One such spatially nonuniform solution close to the transition point
is shown in Fig.~\ref{fig:small-l-solution}. As $\bar{H}$ increases, the optimal solution
turns, for most of the time $0<t<1$, into a stationary ``cnoidal" solution for $p$ which
drives an $h$-profile which is non-uniform in $x$, but is uniformly translating in the vertical direction.
The same solution appears in the problem of the one-point height distribution for the KPZ
equation on a ring~\cite{SMS2018}, and we use it in
Sec.~\ref{sec:small-ell} to calculate the theoretical curves in
Figs.~\ref{fig:l-0125-pi-rf-lbda} and~\ref{fig:small-l-solution},
which match the numerical results quite well.


\begin{figure}
\includegraphics[width = .8\textwidth]{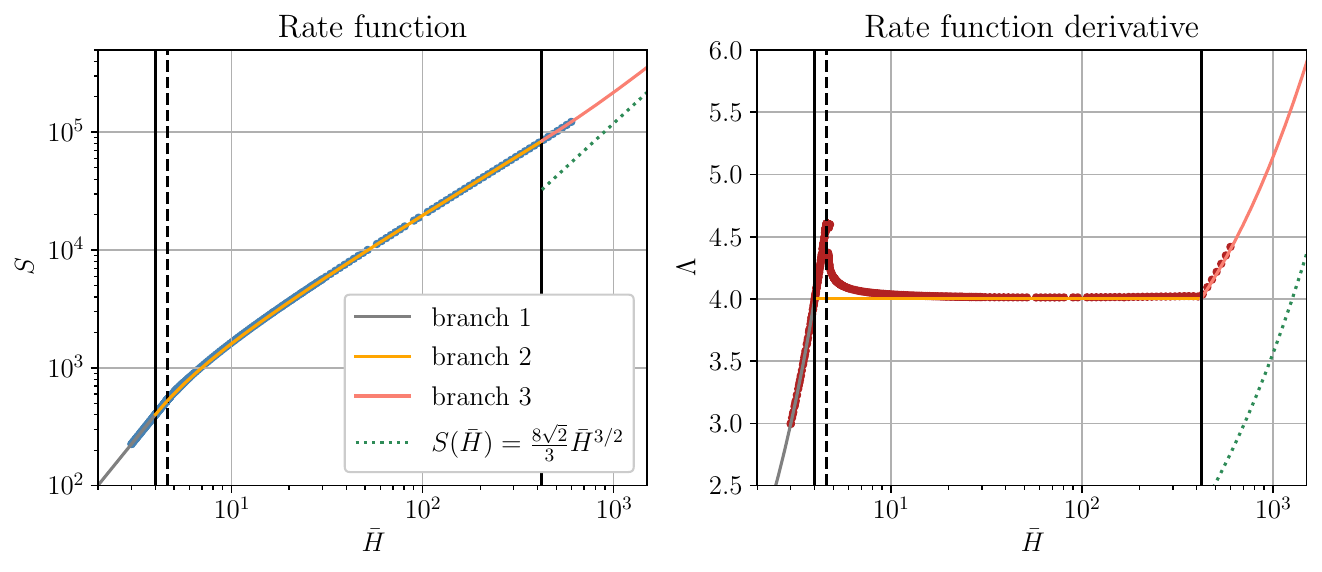}
\caption{Analytical large-$\ell$
results of Sec.~\ref{sec:hL} (lines)
vs. numerics (dots)
for the rate function $S(\bar{H})$ (left) and Lagrange
multiplier $\Lambda(\bar{H})$ (right) for a rescaled system size
of $\ell = 16 \pi$. The
numerical computations were performed at
resolutions $n_x = 1024$ and $n_t = 4000$.
The analytically found branches for the action $S(\bar{H})$
and Lagrange multiplier $\Lambda(\bar{H})$ at large $\ell$ are
drawn as the colored lines according
to the results summarized
in Eq.~(\ref{eq:action-summary}), with~$\Lambda(\bar{H})$
then obtained from Eq.~(\ref{BH092}).
The solid vertical lines indicate the theoretical critical points in
the large-$\ell$ limit at $\bar{H} \simeq 4$ and $\bar{H} \simeq \ell^2 / 6$.
The dashed vertical line shows where
the spatially uniform solution becomes unstable in the $(m = 11)$ spatial
Fourier mode, as given by the minimization of Eq.~(\ref{BH260})
over the allowed wave numbers.
The dotted green line corresponds to the asymptotic behavior
$S(\bar{H}) =(8 \sqrt{2}/3) \bar{H}^{3/2}$, see Eq.~(\ref{hugeHbar}).
A better resolved transition region close to
$\bar{H} = 4$ is shown in
Fig.~\ref{fig:branch2-transition-diff-l}.}
\label{fig:l-16-pi-rf-lbda}
\end{figure}

\begin{figure}
\includegraphics[width = \textwidth]{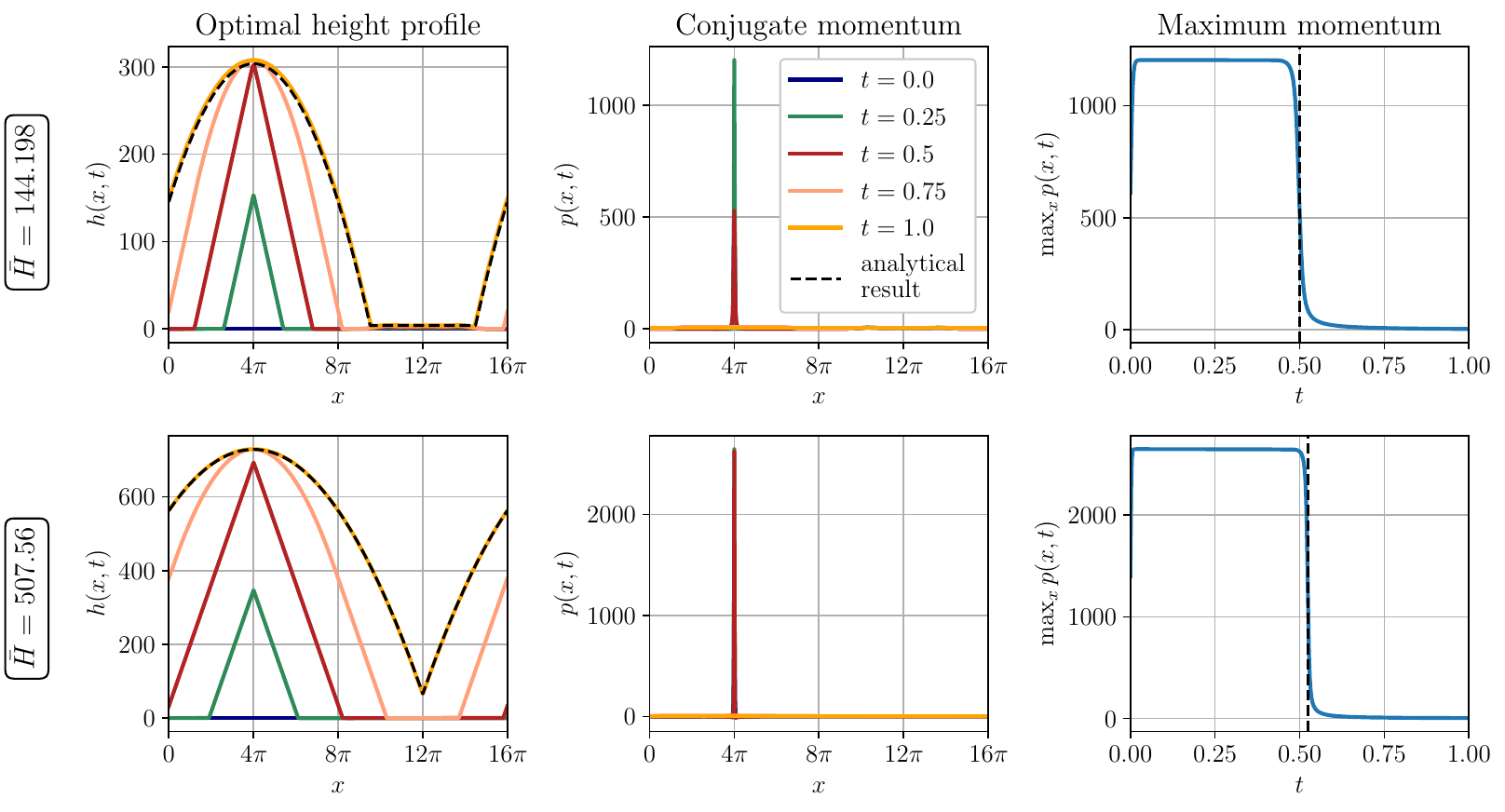}
\caption{Numerically found least-action solutions of Eqs.~(\ref{BH050})
and~(\ref{BH060}) for two mean heights $\bar{H} = 144.198$ (top row)
and $\bar{H} = 507.56$ (bottom row) at $\ell = 16 \pi$ with
numerical resolution $n_x = 2048$ and $n_t = 4000$. Both solutions
are one-soliton solutions for branch 2 and 3, respectively.
The action of the solution in the top row is $S = 28723.766$,
which deviates by $0.4\%$ from the predicted large-$\ell$
result $S = 28590.660$ given by Eq.~(\ref{D190}).
The action for the bottom row solution
is $S = 102754.528$, so it deviates by $0.2\%$ from the
predicted value $S = 102522.498$ for this mean height given by Eq.~(\ref{D300}).
The qualitative difference of these solutions is whether the two edges
of the growing triangle for $h$ collide. The left
column shows the optimal height profile $h(x,t)$ at different
times $t$, together with the prediction~(\ref{D100}),
(\ref{uniform}) and~(\ref{D290}) for the final time
$t = 1$. The center column shows the corresponding conjugate
momentum density $p(x,t)$. The right column shows the spatial
maximum $\max_x p(x,t)$ over time to visualize the lifetime of
the stationary soliton solution and compare it to the analytical
expressions $\tau = 1/2$ for branch 2 and Eq.~(\ref{tauHbar}) for branch
3 (dashed vertical lines).}
\label{fig:branch-2-3-solutions}
\end{figure}

Next, we turn to the more complicated and interesting case of large
$\ell$.
For $\ell = 16 \pi$ the minimization of the augmented action~(\ref{eq:action-lbda-mu})
leads to the results for the rate function $S(\bar{H})$ and Lagrange
multiplier $\Lambda(\bar{H})$ shown
in Fig.~\ref{fig:l-16-pi-rf-lbda}. In addition to branch 1 we observe two other branches of solutions.
Branch 2 is observed to the right of a narrow
transition region close to $\bar{H} \simeq 4$. On this branch  the action $S(\bar{H})$ is
approximately a linear function, while  $\Lambda$ is almost constant. Further, for much larger $\bar{H}$,
there is a smoothed-out second-order transition from branch 2 to a
third branch 3 with a different scaling behavior.
The optimal paths for branches 2 and 3 are shown in
Fig.~\ref{fig:branch-2-3-solutions}. They consist of strongly localized large-amplitude stationary
solitons of $p$ that drive an outgoing almost triangular structure of $h$ (or two antishocks
of $V(x,t) = \partial_x h(x,t)$, see Sec.~\ref{sec:hL}. The solution, corresponding to branch 2,
clearly emerges via a subcritical, rather than supercritical bifurcation. Strikingly, the soliton
has a well-defined life time which is very close to $1/2$.  The
difference between branches 2 and 3 is that, for branch 3, the two edges
of the triangular structure of $h(x,t)$ collide before the final time $t=1$ is reached,
while for branch 2 they do not.


These crucial findings will guide our stationary-soliton-based asymptotic theory for large $\ell$ that we develop
in Sec.~\ref{sec:hL}. There we give an analytical description of the optimal paths
for branches 2 and 3, which are the only relevant ones for large
$\ell$. There we establish a first-order transition at $\bar{H} \simeq 4$ for large but finite $\ell$
and show that it becomes ``accidentally" second order in the limit of $\ell \to \infty$.
We also find that the smoothed-out second-order
transition from branch 2 to branch 3 occurs at $\bar{H} = \ell^2 / 6$. The resulting
analytical predictions, indicated by the lines in
Figs.~\ref{fig:l-16-pi-rf-lbda} and ~\ref{fig:branch-2-3-solutions}, are in good agreement with numerics
at large, but finite $\ell$.

\begin{figure}
\includegraphics[width = .6\textwidth]{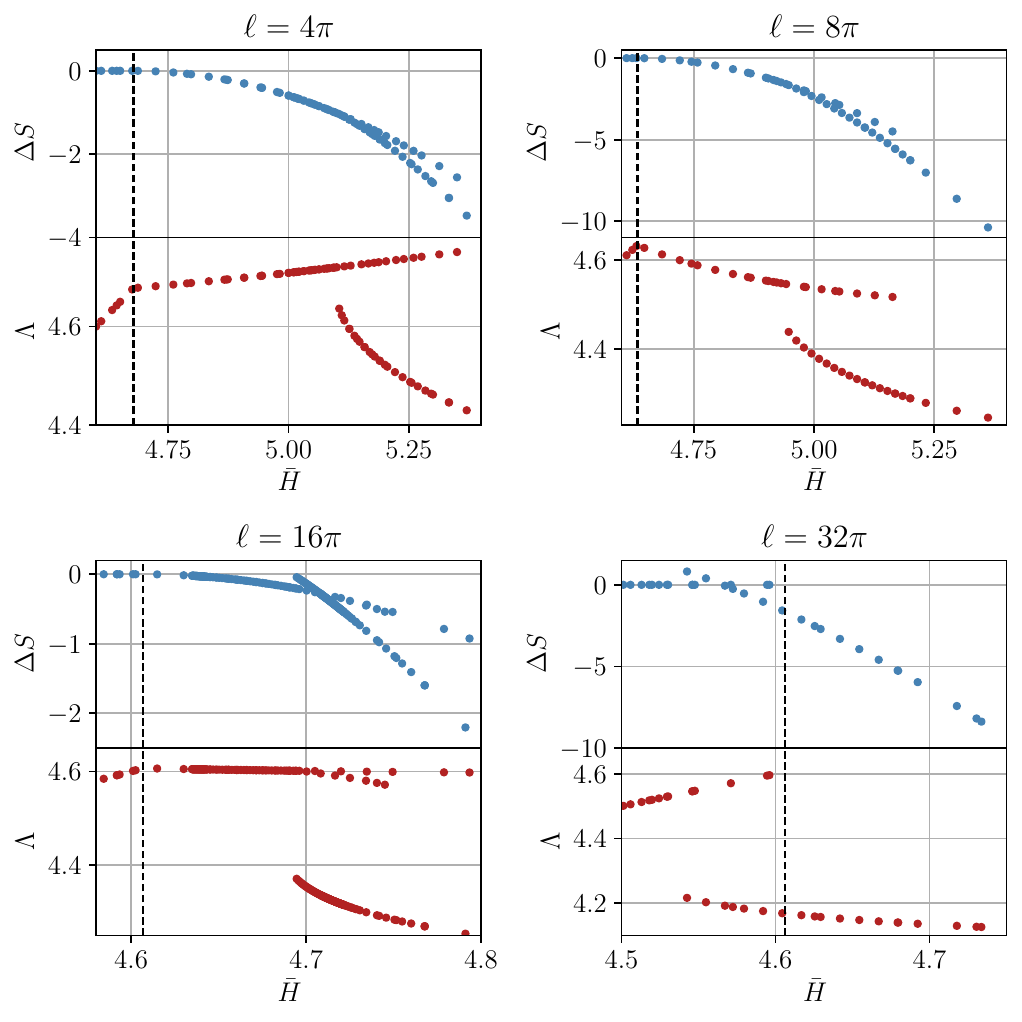}
\caption{Numerically computed action $S(\bar{H})$ and
Lagrange multiplier $\Lambda(\bar{H})$ for
different $\ell$ in the transition region not far from
$\bar{H} = 4$, where the transition between branches 1 and 2  at large $\ell$ is predicted,
see Eq.~(\ref{eq:action-summary}). Using the numerical minimization techniques
described in the Appendix, we search for (possibly multiple distinct local) minimizers
of the action functional~(\ref{eq:action-func}) for given $\ell$ and $\bar{H}$. When there are
more than one minimizers for the same  $\ell$ and $\bar{H}$,
the one with the least action gives the true optimal solution. For a better visualization shown is
the difference between the numerically computed action of the found solutions and the
action $\ell \bar{H}^2/2$ of the spatially
uniform solution. The vertical dashed lines indicate the
smallest $\bar{H}$ where a spatial Fourier mode $q = 2 \pi m / \ell$
of the uniform
solution first becomes unstable according to Eq.~(\ref{BH260})
($m = 3$ for $\ell = 4 \pi$, $m = 5$
for $\ell = 8 \pi$, $m = 11$ for $\ell = 16 \pi$, and $m = 21$
for $\ell = 32 \pi$). For $\ell = 4 \pi$, $8 \pi$
and $16 \pi$, the rate function displays both a
second-order transition at the predicted point, and a first-order transition
at slightly larger $\bar{H}$ where the one-soliton solution (see
the top row of Fig.~\ref{fig:branch-2-3-solutions}), described
theoretically in Sec.~\ref{0p5}, becomes
optimal. At
the largest $\ell = 32 \pi$, only a first-order transition
from the uniform to the one-soliton solution is observed, while
oscillating solutions are irrelevant. See Sec.~\ref{sec:dpt}
for a more detailed analysis of the transition region at large but
finite~$\ell$. Note that for $\ell = 16 \pi$
one can also see, around $\bar{H} = 4.7$, another
oscillating solution with a second superimposed wave number.
Numerical resolutions used: $n_x = 128$ and $n_t = 2000$
for $\ell = 4 \pi$, $n_x = 512$ and  $n_t = 4000$ for $\ell = 8 \pi$,
$n_x = 1024$ and $n_t = 4000$ for $\ell = 16 \pi$, and $n_x = 2048$ and
$n_t = 4000$ for $\ell = 32 \pi$.}
\label{fig:branch2-transition-diff-l}
\end{figure}

At moderate $\ell$ the transition region where the spatially uniform
solution~(\ref{BH210}) of branch 1 becomes sub-optimal is quite
complex, as one can appreciate from
Fig.~\ref{fig:branch2-transition-diff-l}.
We see that, in general, there are both first and second order
transitions in this region: The uniform solution becomes
linearly unstable for some $m > 1$, leading to second-order
transitions, but there is also a competition with the (subcritical) one-soliton
solution. The subcritical scenario clearly wins for sufficiently large $\ell$. Indeed, for $\ell = 32 \pi$
we observe only a first-order
transition from the spatially uniform to the soliton solution,
while the linear instability becomes irrelevant.


\begin{figure}
\includegraphics[width = .66\textwidth]{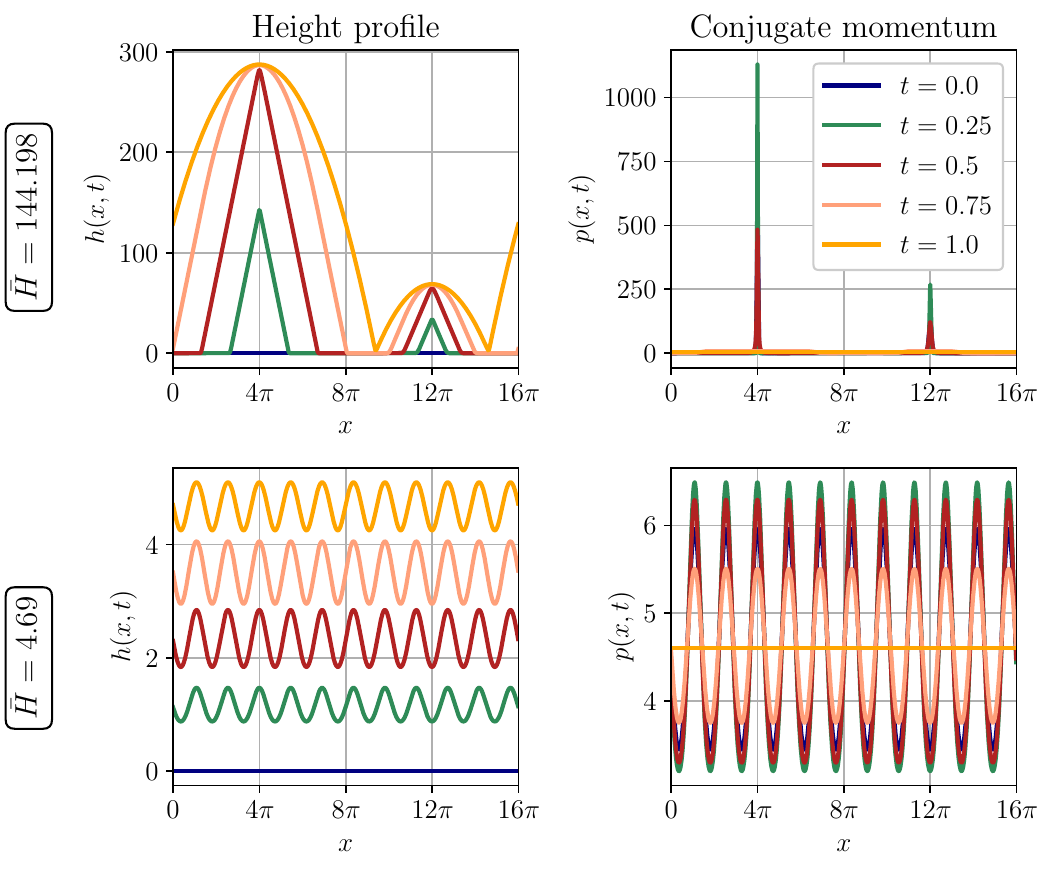}
\caption{Two types of sub-optimal solutions to Eqs.~(\ref{BH050})
and~(\ref{BH060}) for $\ell=16\pi$. The top row (obtained with resolution $n_x = 2048$ and $n_t = 4000$)
shows a
two-soliton solution for the same value of $\bar{H} = 144.198$ as
in the top row of Fig.~\ref{fig:branch-2-3-solutions}. The action
of the two-soliton solution, $S = 28746.682$, is larger than that
for the one-soliton solution for the same $\ell$ and $\bar{H}$.
The bottom row shows a spatially oscillating solution,
originating from linear instability of the $m = 11$ mode of the spatially
uniform solution for $\bar{H} \gtrsim 4.607$ as predicted by Eq.~(\ref{BH260}). Here the
resolution is  $n_x = 1024$ and
$n_t = 4000$.
As can be seen from
Fig.~\ref{fig:branch2-transition-diff-l} (bottom left),
this family of solutions can be optimal for
small $\bar{H}$ and moderate $\ell$,
but it becomes irrelevant as large $\ell$.}
\label{fig:branch-2-3-solutions-subdominant}
\end{figure}

Note that, for branch 2, in addition to stationary single-soliton
solutions of the OFM equation,  discussed so far, there are also stationary multi-soliton solutions
consisting of two or more (almost) non-interacting strongly localized stationary solitons
of $p$ and corresponding expanding triangles of $h$. One such solution, which we observed numerically, is
shown in the top row of
Fig.~\ref{fig:branch-2-3-solutions-subdominant}. We found, however,
that such solutions always have a larger action than
the one-soliton solution for the same $\ell$
and $\bar{H}$. Therefore, the one-soliton solution indeed seems to provide
the optimal solution. In the limit $\ell \to \infty$,
these multi-soliton solutions  -- a soliton gas --  would contribute to the
pre-exponential factor for ${\mathcal P}(\bar{H}, \ell)$, but
pre-exponential factors are beyond the scope of this paper. Additionally, in the
bottom row in Fig.~\ref{fig:branch-2-3-solutions-subdominant},
we show an optimal path for $\ell = 16 \pi$ and close
to $\bar{H} = 4$, which emerges through linear instability of
the $(m = 11)$-mode. Later on, however, it is overtaken by the
one-soliton solution.


\section{Large-$\ell$ asymptotics: Rise and fall of the soliton} \label{sec:hL}
%
\subsection{General description of the solution} \label{sec:hLg}
Guided by our numerical solutions and by the previous works on the one-point KPZ height
statistics on the line~\cite{MKV} and on a ring~\cite{SMS2018}, here we find approximate
asymptotic solutions of Eqs.~(\ref{BH050})-(\ref{BH070}) which give rise to two nontrivial
branches (we call them branches 2 and 3) of the large-deviation function $S(\bar{H})$ for large $\ell$.
As we found, for both branches the maximum one-point height of the interface $H=\max h(x,t=1)$ turns
out to be very large: $H\gg 1$.   Therefore, in addition to the strong inequality $\ell\gg 1$,
we can also use the strong inequality $H\gg 1$. This  allows us to construct ``inviscid"  asymptotic
solutions in different regions of space, separated by discontinuities of proper types. Like their
numerical counterparts, the analytical solutions exhibit two distinct stages in time, with an abrupt
transition between them at some branch-dependent intermediate time $0<t=\tau<1$ which we will determine.

For $0<t<\tau$ the solution has the form of a strongly localized stationary soliton of $p(x,t)$
and ``antishock" of $V(x,t)= -\partial_x h(x,t)$ which were previously identified in the problem
of one-point height statistics on the line~\cite{MKV,KMSparabola} and on a ring~\cite{SMS2018}.
The characteristic width,  $O(1/\sqrt{H})$,  of the soliton-antishock structure is much less than
unity. Outside of the soliton-antishock one has $p(x,t) \simeq 0$. As a result, Eq.~(\ref{BH060})
is obeyed trivially and, at distances $\gtrsim 1$ from the soliton, $h(x,t)$ follows the deterministic KPZ dynamics
\begin{equation}\label{KPZdet}
\partial_th=\partial_x^2h+\frac{1}{2}\left(\partial_xh\right)^2\, ,
\end{equation}
which is equivalent to the Burgers equation
\begin{equation}\label{Burgers}
\partial_tV+ V \partial_x V =\partial_x^2V
\end{equation}
for the field $V(x,t) =-\partial_x h(x,t)$. In addition, the diffusion term in Eq.~(\ref{Burgers})
can be also neglected  at large distances~\cite{MKV}, and one arrives at the inviscid Hopf equation
\begin{equation}\label{D070}
\partial_tV+V\partial_x V=0\,.
\end{equation}
The stationary soliton-antishock structure drives an almost triangular configuration of $h(x,t)$
which is expanding outwards~\cite{MKV}. The height of the triangle grows linearly with time, while
its two edges propagate with a constant speed as ``ordinary" shocks of $V(x,t)$ obeying Eq.~(\ref{Burgers})
or, when treated as discontinuities, obeying Eq.~(\ref{D070})~\cite{MKV}. The positions of these shocks
at $t=1$ determine the boundaries of the  ``impact region" of the soliton-antishock structure. When the
size of the impact region, which scales as $O(\sqrt{H})$~\cite{MKV}, is shorter than the rescaled system
size $\ell$ (this happens when $\bar{H}$ is not too large, see below), there is also an external region
where the uniform solution $p(x,t)=\Lambda =\text{const}$ and $V(x,t)=0$ holds, see Eq.~(\ref{BH210}).
The external uniform solution holds for all times $0<t<1$, and it contributes to the large-deviation
function of $\bar{H}$. In the inviscid limit  the regions of zero and nonzero $p$ are divided by a
stationary discontinuity. This regime corresponds to  branch 2.

Branch 3 appears when, due to the periodicity of the system,  the ordinary shocks of $V(x,t)$
collide with each other before the final time $t=1$ is reached. In this case the impact region
of the soliton-antishock structure extends to the whole system, and a region of the uniform solution does not appear.

For the solution to obey the boundary condition~(\ref{BH070}), the $p$-soliton must turn into a
constant $p= \Lambda$ at $t=1$. Remarkably, as we have seen in our numerical results for large $\ell$,
the soliton rapidly decays in the vicinity of a well-defined time $t=\tau<1$. For both branches 2 and 3,
the subsequent dynamics, at $\tau<t<1$,
gives only a subleading contribution (which we neglect, alongside with other subleading contributions)
to the maximum one-point height $H$ and to the action. This stage is important, however, for determining $\bar{H}$.
We can qualitatively understand this nontrivial temporal structure of the solutions from the viewpoint of action
minimization: First, for $0 \leq t \leq \tau$, the interface is efficiently driven upward by a stationary
$p$-soliton, in the same manner as for the one-point height PDF of the KPZ equation  on the line~\cite{MKV}
and on a ring~\cite{SMS2018}. Then, quickly suppressing the soliton at an intermediate time $0<\tau < 1$ and
evolving the interface according to the almost free KPZ dynamics for $\tau < t \leq 1$ increases considerably
the average height~$\bar{H}$ for a negligible additional cost in terms of action. The optimal value of $\tau$
is the one that minimizes the action for a given $\bar{H}$.

As an overview, we present here the action $S(\bar{H}, \ell)$ at leading order for large $\ell$,
as will be derived in subsections~\ref{0p5} and~\ref{hH}:
\begin{align}
S(\bar{H}, \ell) \simeq \left \lbrace
\begin{array}{lcc}
\tfrac{\bar{H}^2}{2} \ell\,, \quad & -\infty < \bar{H} \leq 4\,, \quad &\text{(branch 1)}\\
\left(4 \bar{H} - 8\right) \ell\,, \quad & 4 < \bar{H} \leq \tfrac{\ell^2}{6}\,,\quad & \text{(branch 2)}\\
\bar{H}^{3/2} \Phi \left(\bar{H} / \ell^2\right)\,, \quad &\tfrac{\ell^2}{6} < \bar{H} < \infty\,, \quad & \text{(branch 3)}
\end{array}
\right.
\label{eq:action-summary}
\end{align}
where the function $\Phi(\dots)$ is defined in Eq.~(\ref{SHblarge}) and
obeys $\Phi(z \to \infty) \to 8 \sqrt{2} /3$. The first line in Eq.~(\ref{eq:action-summary})
comes from the uniform solution~(\ref{BH210}). The first two lines manifestly reveal the large-deviation
scaling~(\ref{eq:pdf-min2}), while the third line does not.

Now we proceed to a more detailed description of the solutions, and we will start with branch 2.

\subsection{Branch 2} \label{0p5}
\begin{figure}
  \centering
 \includegraphics[width=10cm, clip=, bb= 80 60 800 560]{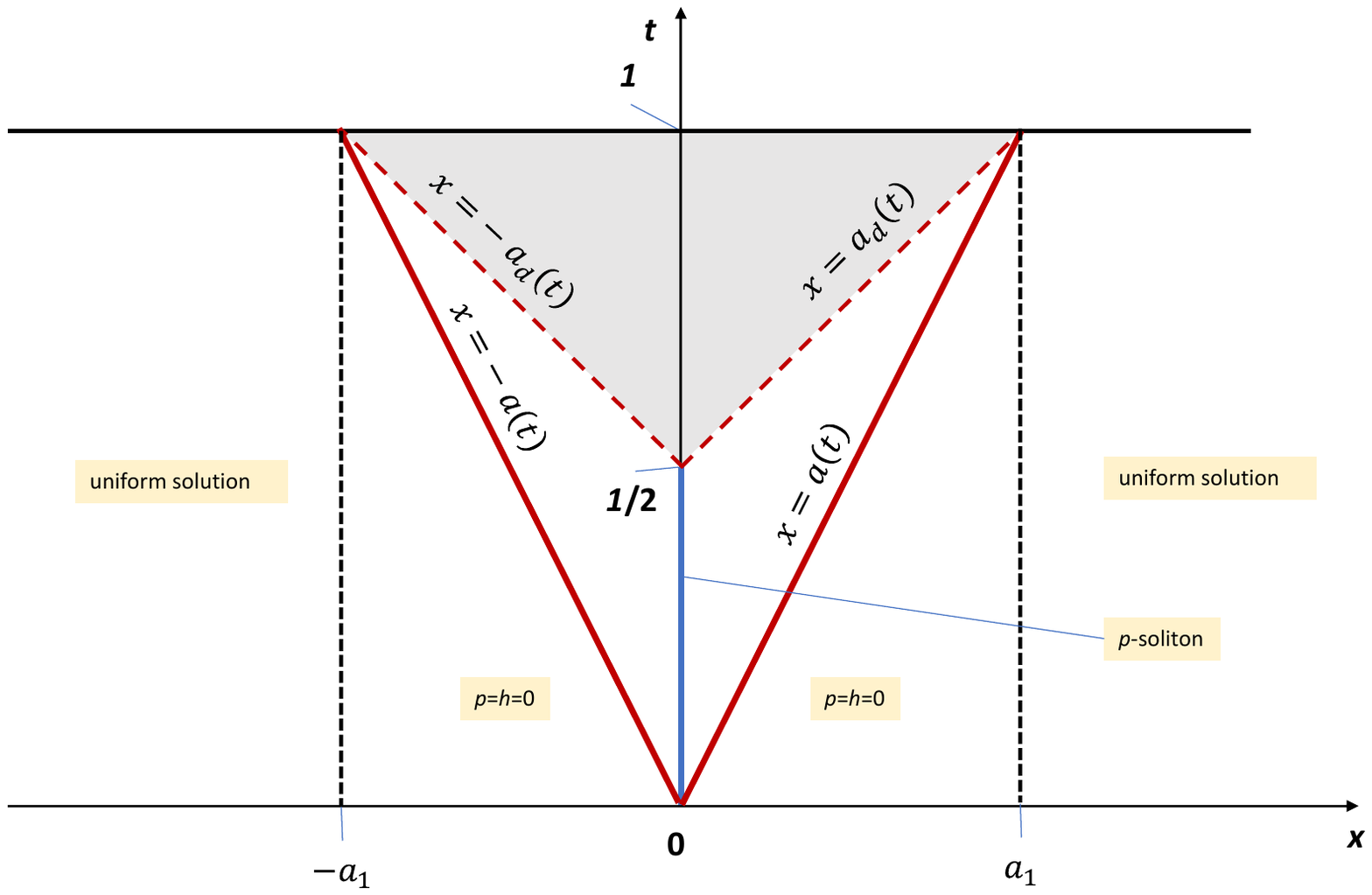}
  \caption{A space-time map of the optimal path of the system which determines the large-deviation
  function $S(\bar{H})$ of branch 2 in the  large-$\ell$ limit.
  The thick blue line shows the position of the stationary $p$-soliton and $V$-antishock. The thick red
  lines show the positions of propagating ordinary shocks, see Eq.~(\ref{D020}). The dashed red lines
  show the positions of propagating weak discontinuities caused by the rapid decay of the $p$-soliton at $t=\tau=1/2$,
  see Eq.~(\ref{D082}). The  weak discontinuities catch up with the (twice as slow) ordinary shocks at $t=1$.
  This condition selects $\tau=1/2$.   The  boundaries of the impact region of the soliton-antishock are shown
  by the dashed black lines. Outside of the impact region, the uniform solution (\ref{uniform}) holds.
  The regions between the dashed black lines and the thick dark lines correspond to the trivial solution~(\ref{D040}).
  In the regions between the dashed and thick red lines, there remain linear $h$-profiles of the
  original ``triangular solution", see Eq.~(\ref{D050}). Inside the gray triangle the solution is described
  by Eqs.~(\ref{D080}) and~(\ref{D160}) leading to Eq.~(\ref{D100}).}
  \label{fig4_0}
\end{figure}

Due to a translational symmetry of the problem~(\ref{BH050})-(\ref{BH070}), we can place the soliton-antishock
structure at $x=0$ (see  Fig.~\ref{fig4_0}) so that, to the leading order, $H\simeq h(0,\tau)$.
As explained above, at $H\gg 1$, the $p$-soliton can be considered as a point-like object. We will only need
the value of its ``mass",  $\int dx\, p(x,t)$ which, by virtue of Eq.~(\ref{BH060}), is conserved. Using
the explicit expression for the soliton, $p(x,t)=p_s(x) = 2 c \cosh^{-2} (\sqrt{c/2}\, x)$~\cite{MKV},
where $c=H/\tau$, we obtain
\begin{equation}\label{D010}
\int_{-\infty}^{\infty} dx\, p_s(x) =  \sqrt{\frac{32 H}{\tau}}\,.
\end{equation}
The base of the triangular structure of the $h$-profile is equal to
\begin{equation}\label{D020}
2a(t)=\sqrt{\frac{2H}{\tau}}\, t\,,
\end{equation}
while the triangle's height is
\begin{equation}\label{D030}
h(0,t)=\frac{Ht}{\tau}\,,\quad 0<t<\tau\,.
\end{equation}
Let us denote the total size of the impact region of the soliton-antishock structure
by $2a_1$, where $a_1 \equiv a(t=1)$. In the region  $a(t)<|x|<a_1$ we have
\begin{equation}\label{D040}
p=h=0\,.
\end{equation}
The triangular profile of $h$ on the interval $0<|x|<a(t)$ is described by the expressions~\cite{MKV}
\begin{equation}\label{D050}
p(x,t)=0\,,\quad h(x,t)
=H\left(\frac{t}{\tau}-\frac{\sqrt{2}|x|}{\sqrt{H\tau}}\right)
\,,\quad\text{and}\quad
V(x,t)=-\partial_xh(x,t) = \tilde{V}\, \mbox{sign\,}x\,,
\end{equation}
where
\begin{equation}\label{D060}
\tilde{V}=\sqrt{\frac{2H}{\tau}}\, .
\end{equation}
As one can see from Eqs.~(\ref{D020}) and~(\ref{D060}), the ordinary shocks propagate
with the speed $\tilde{V}/2$, as to be expected from Eq.~(\ref{Burgers}) or~(\ref{D070})~\cite{Whithambook}.


After the rapid decay of the soliton at $t=\tau$, the ``post-soliton" solution (in the region to be determined)
can be described by the ideal hydrodynamic equations corresponding to the inviscid limit of Eqs.~(\ref{BH050})
and~(\ref{BH060}):
\begin{eqnarray}
 \partial_tV +V \partial_xV&=& -\partial_x p\, , \label{BH0501}\\
  \partial_tp+\partial_x\left(pV\right)&=& 0\,. \label{BH0601}
\end{eqnarray}
The $V$-antishock  now plays the role of a discontinuity which undergoes a decay starting from $t=\tau$.
In the leading order we can neglect the $-\partial_x p$ term, so that Eq.~(\ref{BH0501}) becomes the Hopf
equation~(\ref{D070}). Its solution is
\begin{equation}
 V(x,t)=\frac{x}{t-\tau}\,. \label{D080}\\
\end{equation}
Plugging Eq.~(\ref{D080}) into Eq.~(\ref{BH0601}) and using the ``final" condition~(\ref{BH070})
on $p(x,t=1)$, we obtain
\begin{equation}
p(x,t) =\frac{\Lambda(1-\tau)}{(t-\tau)}\,. \label{D160}
\end{equation}
The solution~(\ref{D080}) and~(\ref{D160}) holds at $t>\tau$ and $|x|\leq a_d(t)$. The boundaries of this region,
\begin{equation}\label{D082}
x= \pm a_d(t)\equiv\tilde{V}(t-\tau)\,,
\end{equation}
represent weak discontinuities, moving with the speed $\tilde{V}$ -- that is twice as fast as
the ordinary shocks at $x=\pm a(t)$, see Eq.~(\ref{D020}). Our simulations show
that the weak discontinuities catch up with the shocks at $t=1$.    The corresponding condition can
be written as $a_d(1) = a_1$, and it yields $\tau=1/2$\footnote{\label{footnote:minimumS1}We also
obtained $\tau=1/2$ analytically by solving the problem for a general $\tau$ and then minimizing the
resulting action with respect to $\tau$. These calculations are somewhat cumbersome, and we do not show them here.}

Therefore, during the second stage of the dynamics, $1/2<t<1$, $V(x,t)$ is described by the following expressions:
\begin{equation}\label{D090}
V(|x|\leq a_d(t),t)=\frac{x}{t-1/2}\,,\quad V(a_d(t)\leq|x|\leq a(t),t)=\pm\tilde{V}
\,,\quad V(a(t)<|x|< a_1,t)=0\,.
\end{equation}
Using the relation $V(x,t)=-\partial_x h(x,t)$, we can obtain the $h$-profile at any time $1/2<t<1$
by integrating Eq.~(\ref{D090}) over $x$. The result describes a parabolic profile of $h$ at $|x|<a_d(t)$,
flanked by the linear profiles at $a_d(t)<|x|<a_1$ corresponding to the triangular structure of $h(x,t)$ of
the first stage the dynamics.   At $t=1$ the parabolic profile takes over the whole interval $|x|<a_1$, and we obtain
\begin{equation}\label{D100}
h(x,t=1)=H-x^2\,,\quad |x|<a_1=\sqrt{H}.
\end{equation}
At $|x|>a_1$ the uniform solution holds:
\begin{equation}\label{uniform}
h(|x|>a_1,t)=\Lambda t\, ,\quad p(|x|>a_1,t)=\Lambda\, .
\end{equation}
Now we evaluate the contributions of the uniform solution to the action, $\Delta S_u$, and to the average
height, $\Delta \bar{H}_u$,  at $t=1$. As $\ell$ goes to infinity, we can neglect the difference between the
total system length $\ell$ and the length of the domain of uniform solution $\ell-2a_1$, and obtain
\begin{equation}\label{D114}
\Delta S_u=\Lambda^2\ell/2\,\mbox{~~~and~~~} \Delta \bar{H}_u=\Lambda\,.
\end{equation}
The leading-order contribution of the soliton-antishock solution to the action is~\cite{MKV}
\begin{equation}\label{D140}
\Delta S_s=\frac{8\sqrt{2}}{3}\, \frac{H^{3/2}}{\sqrt{\tau}}=\frac{16 H^{3/2}}{3}\,.
\end{equation}
This contribution comes from the first stage of the process, $0<t<1/2$, while the second stage gives
only a subleading contribution which we neglect.
The second stage, $1/2<t<1$ does contribute to $\bar{H}$, however. Using Eq.~(\ref{D100}), we obtain
\begin{equation}\label{D130}
\Delta\bar{H}_s=\frac{4 H^{3/2}}{3\ell}\,.
\end{equation}
What remains to be done is to determine $\Lambda$, to collect the contributions to $S$ and $\bar{H}$,
and to eliminate $H$ in favor of $\bar{H}$ and $\ell$.
In order to determine $\Lambda$, we use the local conservation of $p(x,t)$ evident in Eq.~(\ref{BH060}).
Because of this local conservation law,
the total soliton ``mass", see Eq.~(\ref{D010}), must be equal to the integral of the solution~(\ref{D160})
for $p(x,t)$ over $x$ from -$a_1$ to $a_1$. This condition yields a remarkably simple result: $\Lambda=4$,
a constant value (up to small subleading corrections).
Combining Eqs.~(\ref{D114})-(\ref{D130}), we obtain
\begin{equation}\label{D180}
\bar{H}=4+\frac{4 H^{3/2}}{3\ell}\,,
\quad
S=8\ell+\frac{16 H^{3/2}}{3}\, .
\end{equation}
Eliminating $H$, we arrive at the leading-order result for the large-deviation function of $\bar{H}$
for branch 2 in the limit of large $\ell$, which was announced in the second line of Eq.~(\ref{eq:action-summary}):
\begin{equation}\label{D190}
S=\left(4\bar{H} -8\right)\, \ell\,.
\end{equation}
This expression obeys the large-deviation scaling~(\ref{BH120}). As was to be expected, the actions
of branch 1 and 2 coincide at
$\bar{H}=\bar{H}_{\text{c}}=4$. Noticeably, their first derivatives with respect to $\bar{H}$
also coincide at this point.
In addition, using Eq.~(\ref{BH092}), we see that Eq.~(\ref{D190}) is consistent with $\Lambda=4$,
independently of $\bar{H}$, for branch 2.
We will look into these peculiarities more carefully in Sec.~\ref{sec:dpt}.

One applicability condition of Eq.~(\ref{D190}) is the strong inequality $H\gg 1$.
Using the first relation in Eq.~(\ref{D180}),
we can rewrite this strong inequality in terms of $\bar{H}$ and $\ell \gg 1$:
\begin{equation}\label{Hbarcond}
\bar{H}-4 \gg 1/\ell\,.
\end{equation}
This condition limits $\bar{H}$ from below. A condition on $\bar{H}$ from above distinguishes
branch 2 from branch 3. It demands  that the ordinary shocks of $V(x,t)$ do not collide with
each other until $t=1$\footnote{\label{footnote:unnecessary}While deriving Eq.~(\ref{D114}) we
demanded a \emph{strong} inequality $2\sqrt{H}\ll\ell$. However, when $\bar{H}\gg 1$, the main contribution
to $S$ and $\bar{H}$ comes from the soliton-antishock solution, rather than from the uniform one. As a
result, the strong inequality $2\sqrt{H}\ll\ell$  becomes unnecessary, and a simple inequality suffices.}.
This condition can be written as $2\sqrt{H}<\ell$ or, using Eq.~(\ref{D180}),
\begin{equation}\label{D200}
\bar{H}-4<\frac{\ell^2}{6} \mbox{~~~at~~~}\ell\gg1\, .
\end{equation}
Now we proceed to a description of branch 3.


\subsection{Branch 3} \label{hH}
\begin{figure}
 \centering
 \includegraphics[width=9cm, clip=, bb= 160 60 760 560]{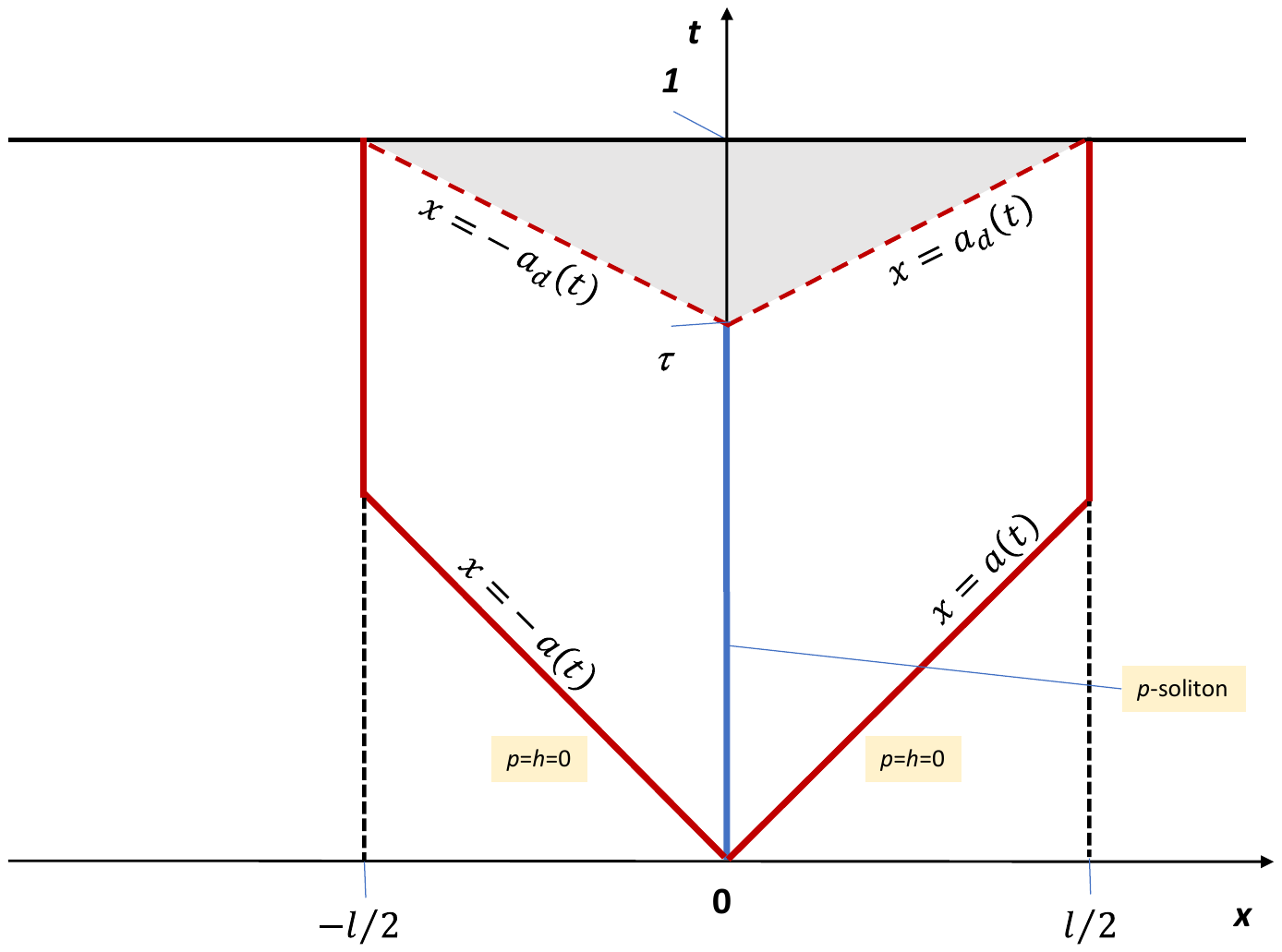}
  \caption{A space-time map of the optimal path of the system which determines the large-deviation
  function $S(\bar{H})$ of branch 3 in the large-$\ell$ limit.
  The notations are similar to those of Fig.~\ref{fig4_0}.
}
\label{fig4_5}
\end{figure}

When the inequality~(\ref{D200}) is violated, the two outgoing ordinary shocks of $V(x,t)$ collide
with each other and merge at $x=\pm \ell / 2$ (which is the same point of the ring) at some $t<1$.
Upon the merger, a single stationary shock appears,  see Fig.~\ref{fig4_5}. Now the impact region of
the soliton-antishock is the whole system: $2a_1=\ell$, and the external region of the uniform solution,
characteristic of branch 2, does not appear here.

Most of the general formulas, derived in the context of branch 2, remain valid for branch 3.
In particular, here too $\tau$ is determined by the condition that the weak discontinuities catch
up with the ordinary shocks at $t=1$. The only difference is that $a_1=\ell/2$ now. Solving the
equation $a_d(1) = a_1$, or
\begin{equation}\label{tau3eq}
\sqrt{\frac{2H}{\tau}} \left(1-\tau\right) = \frac{\ell}{2}\,,
\end{equation}
we obtain
\begin{equation}\label{tau3}
\tau =1+\frac{\ell^2}{16 H}-\frac{\ell\sqrt{\ell^2+32H}}{16 H}\,,
\end{equation}
so that $\tau$ depends on $H$ and $\ell$. Unsurprisingly, Eq.~(\ref{tau3}) yields $\tau=1/2$ in
the boundary case $H=\ell^2/4$, when the size $2a_1$ of the impact region of the soliton-antishock
in an infinite system is equal to the system size $\ell$. When $H$ goes to infinity, $\tau$ approaches $1$.

\begin{figure}
 \includegraphics[width=.4\textwidth]{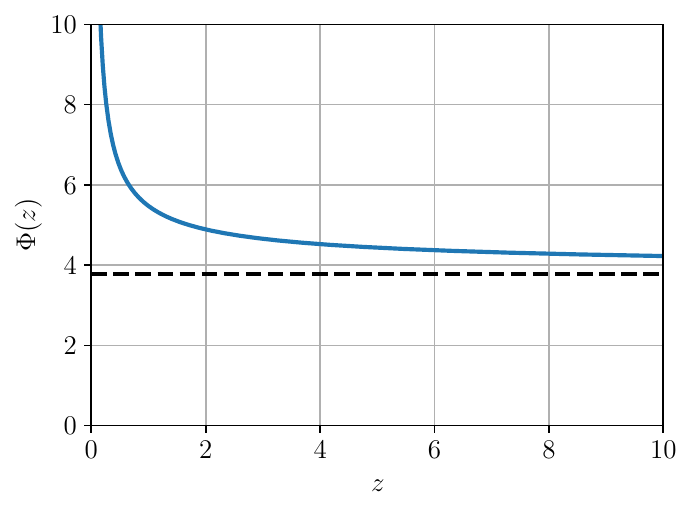}
  \caption{A plot of the function $\Phi(z)$ (blue) which enters  Eq.~(\ref{SHblarge})  for the
  large-deviation function $S(\bar{H},\ell)$ for branch 3 in the large $\ell$ limit. The dashed
  line shows the large-$z$ asymptote $\Phi(z\to \infty)=8\sqrt{2}/3$.}
\label{Phiz}
\end{figure}

We will not repeat here all expressions for $h(x,t)$, $V(x,t)$ and $p(x,t)$ in different regions,
and present only the expression for $h(x,1)$:
\begin{equation}\label{D290}
h(x,1)=H-\frac{x^2}{2(1-\tau)}\,,
\end{equation}
with $\tau$ from Eq.~(\ref{tau3}).
Using this expression, we can evaluate $\bar{H}$. The action $S$ remains the same as in the
first equality in Eq.~(\ref{D140}), and we obtain
\begin{equation}\label{D300}
\bar{H}=H-\frac{1}{24}\, \frac{\ell^2}{(1-\tau)}\,,
\quad
S=\frac{8\sqrt{2}}{3}\, \frac{H^{3/2}}{\sqrt{\tau}}\,.
\end{equation}
Eliminating $H$ from these relations and using Eq.~(\ref{tau3}), we arrive at a leading-order
result for the large-deviation function $S(\bar{H},\ell)$ in the limit of large $\ell$ and very
large $\bar{H}$, which was announced in the third line of Eq.~(\ref{eq:action-summary}):
\begin{equation}
S(\bar{H},\ell) = \bar{H}^{3/2} \Phi\left(\frac{\bar{H}}{\ell^2}\right)\,,\quad \text{where}\quad
 \Phi(z) =\frac{2 \sqrt{2} \,(9 z+1+\sqrt{18z+1})^{1/2} \left(36 z+1+\sqrt{18z+1}\right)}{81 z^{3/2}} \,.
 \label{SHblarge}
\end{equation}
In terms of $\bar{H}$, the condition $H>\ell^2/4$ becomes, in the leading order, $\bar{H}>\ell^2/6$.
As a result, the function $\Phi(z)$ is defined for $z\geq 1/6$, and $\Phi(1/6) = 4 \sqrt{6}$.
A graph of $\Phi(z)$ is depicted in Fig.~\ref{Phiz}.


In the limit of $\bar{H}\gg \ell^2\gg 1$ Eq.~(\ref{SHblarge}) yields
\begin{equation}\label{hugeHbar}
S=\frac{8\sqrt{2}}{3}\bar{H}^{3/2}+\frac{4}{3}\bar{H} \ell+ \dots\,.
\end{equation}
The leading-order term of this expression coincides with the action for a single-point height $H$~\cite{MKV}.
This is to be expected, because for very large $\bar{H}$, $\tau$ approaches $1$, and the difference
between $\bar{H}$ and $H$ becomes relatively small.

The expressions in Eqs.~(\ref{D190}) and~(\ref{SHblarge}) match in the leading order in $\ell$
at the boundary $\bar{H}\simeq \ell^2/6$ between the branches 2 and 3, both giving $(2/3) \ell^3+O(\ell)$.

For completeness, we also present the optimal transition time $\tau$ in Eq.~(\ref{tau3}) in terms of $\bar{H}$ and $\ell$:
\begin{equation}\label{tauHbar}
\tau(\bar{H},\ell)=1+\frac{\ell^2}{12 \bar{H}}-\frac{\ell\sqrt{\ell^2+18
  \bar{H}}}{12 \bar{H}}\,.
\end{equation}


\subsection{Dynamical phase transition} \label{sec:dpt}
In this subsection we resolve the nature of the DPT between
branches~1 and~2, which corresponds to the subcritical bifurcation from the uniform solution~(\ref{BH210})
to the leading-order soliton solution discussed in Sec.~\ref{0p5}. To this end we will have to focus
on subleading corrections that we have previously ignored. We will also present the large-deviation
scaling of $\mathcal{P}(\bar{H},L,T)$ in the limit of~$T \to 0$ at fixed $L$,  in the physical units.

As we have already noticed, the actions $S_1(\bar{H}, \ell)$ and $S_2(\bar{H}, \ell)$, described
by the first and second lines of Eq.~(\ref{eq:action-summary}),
coincide at $\bar{H}=\bar{H}_{\mathrm{c}}=4$ together with their first derivatives $\partial S_1(\bar{H}, \ell) /
\partial \bar{H}$ and $\partial S_2(\bar{H}, \ell)/\partial \bar{H}$
at~$\bar{H}_{\mathrm{c}}=4$. It would be incorrect, however,
to conclude from here that the DPT between branches~1 and~2 at~$\bar{H}=\bar{H}_{\mathrm{c}}$
is of second order. Indeed, the supercritical first bifurcation of the uniform solution~(\ref{BH210})
to a solution with a single maximum of~$h(x,1)$ -- the one with $q = 2 \pi / \ell$
in Eq.~(\ref{BH250}) -- actually occurs, as~$\ell\to\infty$, at much
larger~$\bar{H} \simeq \ell^2 / 16 \gg 4$. Furthermore,
as follows from numerical minimization of Eq.~(\ref{BH250}), instability
of \textit{any} Fourier mode around the uniform solution can only occur
at~$\bar{H} \simeq 4.60334$ (for~$q \simeq 1.34336$). It
is not surprising, therefore, that
at large but finite~$\ell$, and at a slightly shifted transition
point~$\bar{H}_{\mathrm{c}}> 4$ where the actions of branches~1 and~2
are equal, the optimal paths~$h(x,t)$ for branches~1 and~2, that we found numerically,
are dramatically different, and their respective Lagrange
multipliers~$\Lambda$ are not equal. The latter fact means, by
virtue of Eq.~(\ref{BH092}), that at large~$\ell$  we actually observe a first-order DPT, not a second-order one.

To make sense of these facts, we recall that Eq.~(\ref{D190})
for the action of branch 2 is merely a leading order asymptotic
at $\ell \to \infty$. Subleading terms, so far unaccounted for, should remove
the degeneracy of the leading-order results by breaking the accidental continuity
of the first derivative $\partial S(\bar{H}, \ell)/\partial \bar{H}$
at $\bar{H}=\bar{H}_{\mathrm{c}}$, and
rendering the corresponding bifurcation subcritical and the corresponding DPT
first-order. The subleading terms should also account for a slight shift of the critical
point $\bar{H}_{\mathrm{c}}$ to the right from its leading-order
value $\bar{H}_{\mathrm{c}}=4$, as observed in our numerics.

Motivated by the large-$H$ asymptotic of the upper tail of the exact
short-time probability distribution of the one-point height $h(x = 0,t = 1)=H$
on the line, determined in Ref.~\cite{SmithMeerson2018}, we can conjecture the following
subleading terms of $S_2(\bar{H},\ell)$ at large $\ell$:
\begin{equation}\label{D416}
S_2(\bar{H},\ell)=\left(4\bar{H} -8\right) \ell+B H^{1/2}+C H^{-1/2}+\dots\,,
\end{equation}
where $B>0$ and $C$ are numerical constants $O(1)$, which are independent
of $\ell$. The condition $B>0$ is necessary for the equation
\begin{equation}\label{D400}
S_1 \left( \bar{H}_{\mathrm{c}},\ell \right) =
S_2 \left( \bar{H}_{\mathrm{c}},\ell \right)
\end{equation}
to have a solution for $\bar{H}_c$ close to
$4$ at large $\ell$.

\begin{figure}
  \centering
 \includegraphics[width=.4\textwidth]{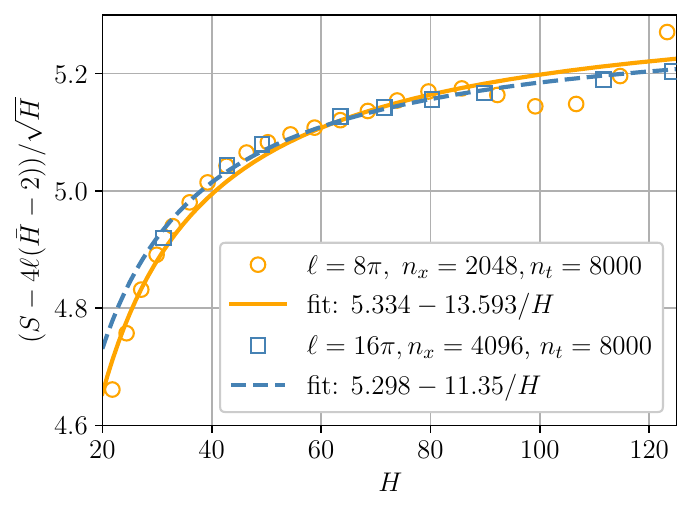}
  \caption{Numerical data from high-resolution optimal path
   computations at rescaled system sizes $\ell = 8 \pi$ and $16 \pi$.
   Assuming an asymptotic expansion~(\ref{D416}), the ordinate should be
   given by the subleading terms $B + C / H$ with constants $B$ and $C$
   that are independent of $\ell$ and $H$. Accordingly, we performed
   least-squares fit with this functional form to the data for $20 \leq H
   \leq 125$. We observe a fair agreement for
   the two rescaled system sizes and determine $B \simeq 5.3$. For the smaller domain
   size $\ell = 8 \pi$ there are small oscillations
   which may come from sub-subleading terms not included in
  Eq.~(\ref{D416}).
}
\label{fig_num}
\end{figure}

To verify Eq.~(\ref{D416}), we plotted in Fig.~\ref{fig_num} our large-$\ell$ numerical results for
$\left[S_2(\bar{H},\ell) - \left(4\bar{H} -8\right)
\ell\right]/\sqrt{H}$ versus $H$. A fair plateau at large $H$ is observed, with $B \simeq 5.3 > 0$ found by fitting.
Now, keeping the first subleading term in Eq.~(\ref{D416})
and the leading-order dependence of $H$ on $\bar{H}$ in Eq.~(\ref{D180}),
we can rewrite Eq.~(\ref{D416}) in terms of $\bar{H}$ and $\ell$:
\begin{equation}\label{D420}
S_2(\bar{H},\ell)=8\ell+4\left(\bar{H} -4\right) \ell
+ \left(\frac{3}{4}\right)^{1/3} B \left[\left(\bar{H}-4\right)\ell\right]^{1/3}
+ \dots \,,
\qquad (\bar{H}-4)\ell\gg 1\,.
\end{equation}
Now Eq.~(\ref{D400}) for the critical point becomes
\begin{equation}\label{D430}
\frac{1}{2} \left(\bar{H}_{\mathrm{c}}-4 \right)^2\ell
= \left(\frac{3}{4}\right)^{1/3} B \left[ \left(\bar{H}_{\mathrm{c}}
-4 \right)\ell\right]^{1/3}+\dots\,,
\end{equation}
Its approximate solution,
\begin{equation}\label{D440}
\bar{H}_{\mathrm{c}} = 4 + 6^{1/5} B^{3/5}\, \ell^{-2/5}+\dots\,,
\end{equation}
describes a small $\ell$-dependent positive shift of the critical point from the leading-order value $4$.
This~$\bar{H}_{\mathrm{c}}$ corresponds to
\begin{equation}\label{D450}
H = \left(\frac{9}{8}\right)^{2/5} B^{2/5} \ell^{2/5} +\dots
\end{equation}
of the branch-2 solution at the critical point. We observe that, for this solution, $H \to \infty$
as $\ell\to \infty$, guaranteeing applicability of our theory at large $\ell$. Going back to the
large-deviation scaling~(\ref{BH120}), we notice that there is now a small but finite jump  $\sim\ell^{-2/5}$
of the derivative $\ell^{-1} \partial S/\partial\bar{H}$ of the effective rate function at the shifted critical
point.  The transition between branches 1 and 2, therefore, is of first order at large but finite $\ell$. Such transitions -- with a finite but small jump of the first derivative of the free energy (or the action) are usually called weakly first order transitions~\cite{Binder}.

By virtue of Eq.~(\ref{BH092}), the subleading correction in Eq.~(\ref{D420}) also removes the degeneracy
of the leading-order result $\Lambda=4$ by adding to it a small $\ell$-dependent correction that goes
to zero as $\ell\to \infty$.

\begin{figure}
  \centering
 \includegraphics[width=.4\textwidth]{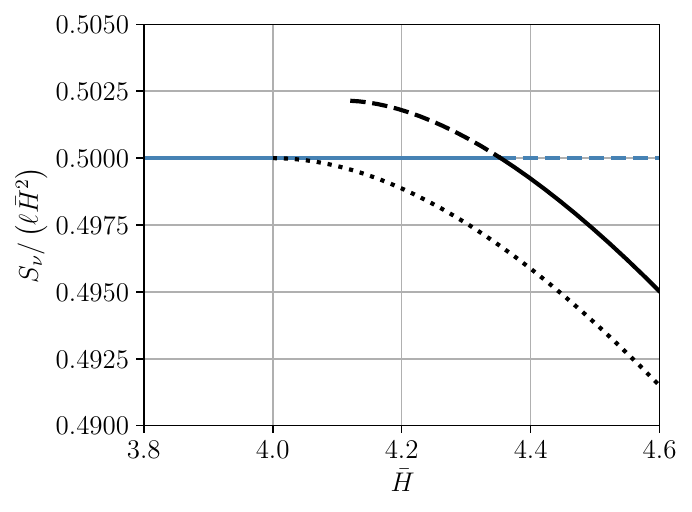}
  \caption{The rescaled actions $S_\nu/(
  \ell \bar{H}^2)$ for $\nu = 1$  and $\nu=2$  versus $\bar{H}$ for a very long domain $\ell = 128\pi$.
 The result for $\nu=1$ (branch~1) is shown by the blue lines, the prediction of  Eq.~(\ref{D420})
 for $\nu=2$ (branch~2) is shown by the black lines. The solid lines correspond to the global minimum of
  the action, whereas the dashed lines correspond to local but not global
  minima of the action. The thin dotted line shows the asymptotic theoretical prediction
for $S_2/ (\ell \bar{H}^2)= (4 \bar{H}-8)\bar{H}^{-2}$ for $\ell \to \infty$.}
\label{fig_cross}
\end{figure}

Using Eq.~(\ref{D420}), we plotted in Fig.~\ref{fig_cross} the actions of branches~1 and~2, normalized
by $\ell \bar{H}^2$, in the
vicinity of the $\bar{H} = \bar{H}_{\text{c}}$. It is clearly seen that the subleading correction removes the degeneracy
and makes the DPT first-order. Furthermore,
the predicted $\bar{H}_{\mathrm{c}}$ from Eq.~(\ref{D440})
for $\ell = 32 \pi$, which is $\bar{H}_{\mathrm{c}}
\simeq 4.6$, is close to our numerical result  $\bar{H}_{\mathrm{c}} \simeq 4.57.$ for this $\ell$, see
Fig.~\ref{fig:branch2-transition-diff-l}.


Note that our arguments in favor of the expansion~(\ref{D416}) are far from rigorous.
In particular, we cannot exclude a very
slow (for example, logarithmic) dependence of the coefficient~$B$ on $H$ in Eq.~(\ref{D416})
based only on the numerical evidence. However,
our main conclusion about the first-order DPT between branches 2 and 3
seems robust.

To conclude this section, we present our large-deviation results, described by the first two lines
of Eq.~(\ref{eq:action-summary}), in
the physical units. Recall that, by taking the
limit $T \to 0$ at fixed  $L$,
we have both $\varepsilon \propto T^{1/2} \to 0$ and $\ell\to \infty$. In this limit only the first
two lines of Eq.~(\ref{eq:action-summary}) are relevant, and we
obtain\footnote{\label{footnote:T}Note the factor of $T$ instead of the customary weak-noise
factor $T^{1/2}$ on the left-hand side
of Eq.~(\ref{D460}).}
\begin{equation}\label{D460}
-\lim\limits_{T\to 0} T\, \ln P(\bar{H},L,T)
=\frac{\nu^2}{D\lambda^2}\,L\, f\left(\frac{\lambda\bar{H}}{\nu}\right)\,,
\mbox{~~~where~~~}
f(w)=\left\{
\begin{array}{lcr}
w^2/2&\mbox{~for~}& w<4\,,\\
4w-8&\mbox{~for~}& w>4\,.
\end{array}
\right.
\end{equation}
As we
elaborated in this subsection, the DPT
in Eq.~(\ref{D460}) at $w = 4$ can be called an ``accidental''
second order DPT in the sense that the optimal paths, that are responsible for the two branches in Eq.~(\ref{D460}),
transition into each other discontinuously, and that the differentiability of the rate function
at the critical point emerges only in
the limit $T \to 0$ at fixed $L$.


\section{Small-$\ell$ asymptotics} \label{sec:small-ell}
We found that our numerical results on the second-order DPT at small $\ell$, shown in Figs.~\ref{fig:l-0125-pi-rf-lbda}
and~\ref{fig:small-l-solution} and described in Sec.~\ref{sec:num},
can be understood in terms of a small-$\ell$ asymptotic solution of the OFM equations~(\ref{BH050})
and~(\ref{BH060}) which was previously found in the context of  the one-point
height distribution on a ring~\cite{SMS2018}. In this solution
the interface is driven by a stationary $\text{dn}^2$  profile (see below) of $p$. The solution represents a finite-amplitude
generalization of a weak sinusolidal modulation with $m = 1$ which results from the second-order DPT from
the uniform solution. This solution is given by the following expressions\footnote{\label{footnote:narrow}This
solution is invalid inside
narrow boundary layers in time at $t=0$ and $t=1$, but their contribution to the action is negligible.}
\begin{align}
h(x,t) &\simeq H t + 2 \ln \text{dn} \left[\frac{2 K(k) x}{\ell},
k \right]\,, \label{eq:h-ellip}\\
p(x,t) &\simeq p_0(x) = \left[\frac{4 K(k)}{\ell} \right]^2
\text{dn}^2 \left[\frac{2 K(k) x}{\ell} , k\right]\,, \label{eq:p-ellip}
\end{align}
where $K(k)$ is the complete elliptic integral of the first kind
and $\text{dn}(\dots)$ is one of the Jacobi elliptic functions~\cite{Jacobi}.
The elliptic modulus $k \in (0,1)$ is determined by $H$ via the relation
\begin{equation}
\frac{8 \left(2 - k^2\right) K^2(k)}{\ell^2} = H\,,
\label{eq:m-determ}
\end{equation}
The action of this solution as a function of $k$ is~\cite{SMS2018}
\begin{equation}
S(k) = \frac{128}{3 \ell^3} K^3(k) \left[2\left(2-k^2\right) E(k)
- \left(1-k^2\right) K(k) \right]\,.
\label{eq:ellip-action}
\end{equation}

At given $\ell\ll 1$, Eqs.~(\ref{eq:m-determ}) and~(\ref{eq:ellip-action}) determine $S$ as a
function of $H$ in a parametric form. The critical point  $\bar{H} = (2 \pi / \ell)^2$ corresponds
to $k=0$, when Eqs.~(\ref{eq:m-determ}) and~(\ref{eq:ellip-action}) reduce to the uniform solution. $k>0$
correspond to supercritical solutions.

In order to recast this dependence in terms of $S(\bar{H},\ell)$,
we need to express $H$ through $\bar{H}$ and $\ell$. Although Eq.~(\ref{eq:h-ellip}) is formally inapplicable
at $t=1$, asymptotically as $\ell \to 0$ we still have
\begin{equation}
\label{Hbarshort}
H - \bar{H}\simeq -\frac{1}{\ell}  \int_{-\ell /2}^{\ell / 2}
2 \ln \text{dn} \left[\frac{2 K(k) x}{\ell},
k \right] dx= \frac{1}{2} \ln \frac{1}{1 - k^2}\,.
\end{equation}
where we have used a product formula for $\text{dn}$~\cite{Wolf}.
Using Eqs.~(\ref{eq:m-determ}) and~(\ref{Hbarshort}), we obtain
\begin{equation}\label{Hbarvsk}
\bar{H}(k) = \frac{8 \left(2 - k^2\right) K^2(k)}{\ell^2}-\frac{1}{2} \ln \frac{1}{1-k^2}\,.
\end{equation}
Equations~(\ref{Hbarvsk}) and (\ref{eq:ellip-action}) determine $S=S(\bar{H},\ell)$ and were
used in Fig.~\ref{fig:l-0125-pi-rf-lbda} to draw the theoretical curves for the action and
Lagrange multiplier (via Eq.~(\ref{BH092}))
at $\ell = \pi / 8$, which agree very well with the numerical action minimization results. Also shown is the
asymptotic action
\begin{align}
S(\bar{H}) \simeq \frac{8 \sqrt{2}}{3} \bar{H}^{3/2}
\label{eq:action-asymp}
\end{align}
as $\bar{H}\to \infty$, which agrees with Eq.~(\ref{hugeHbar}) and can be obtained from
Eqs.~(\ref{eq:ellip-action}) and~(\ref{Hbarvsk}) by considering the limit $k \to 1$
with $E(k) \to 1$ and $K(k) \simeq \tfrac{1}{2} \ln \tfrac{1}{1-k}$. As one can see from
Fig.~\ref{fig:small-l-solution}, the asymptotic relation~(\ref{Hbarshort})
is not yet satisfied for the moderately small $\ell = \pi / 8$: noticeably, the solution $h(x,1)$
at the final time deviates from Eq.~(\ref{eq:h-ellip}). However, the numerically found action
is already accurately described by Eqs.~(\ref{eq:ellip-action}) and~(\ref{Hbarvsk}), because
the difference between~$H$ and~$\bar{H}$ is always subleading -- at most $O(\sqrt{H})$ -- at small $\ell$.


\section{Summary and discussion} \label{summary}
We applied the OFM to evaluate analytically and numerically the short-time PDF $P (\bar{H}, L, t=T)$,
and the optimal paths which dominate this PDF, of the KPZ interface on a ring.  The short-time PDF has
the scaling form~(\ref{eq:pdf-min1}), where $\varepsilon \sim T^{1/2}$ plays the role of the weak-noise
parameter.  The phase diagram of the system
represents the $(\bar{H}, \ell=L/\sqrt{\nu T})$ plane. We were especially interested in the DPTs that occur
in this system at sufficiently large positive $\lambda \bar{H}>0$. We found that, depending on $\ell$, these
DPTs occur via either a supercritical, or a subcritical bifurcation of the ``trivial" (uniform in space)
optimal path of the KPZ interface. The supercritical bifurcations dominate at very small $\ell$, the subcritical
bifurcations dominate at very large $\ell$. In these two limits we obtained asymptotic analytical solutions
for the optimal paths of the system, evaluated the resulting action, and verified the analytical results
numerically. We also found that, as $T$ goes to zero at constant $L$, the PDF acquire a simple large-deviation
form~(\ref{eq:pdf-min2}) and ~(\ref{D460}). Interestingly, the rate function $f(\bar{H})$ exhibits,  at a critical value
of $\bar{H}=\bar{H}_{\text{c}}(\ell)$, a DPT which is \emph{accidentally} second-order.

In the (much more complicated) region of intermediate $\ell=O(1)$ we observed numerically both supercritical,
and subcritical bifurcations of the uniform solution. This region of the phase diagram is presently out of
reach of analytical theory. It would be very interesting, but challenging, to determine the complete phase
diagram of the system in this region. In particular, it would be interesting to locate, somewhere
between $\ell=16 \pi$ and $\ell = 32\pi$, at least one critical point $(\bar{H}_*, \ell_*)$ where the
second order DPT curve $\bar{H}_c^{(2)}(\ell)$ ends when it meets the first order DPT curve $\bar{H}_c^{(1)}(\ell)$.

These tasks will become more feasible if this problem, as described by Eqs.~(\ref{BH050})-(\ref{eq:final-constr}),
joins the list of similar
large-deviation OFM problems for the KPZ equation which have been solved exactly by the inverse scattering
method~(ISM)~\cite{KLD2021,KLD2022}. Indeed, as was previously found in Ref.~\cite{Janas2016},
a canonical Hopf--Cole transformation brings Eqs.~(\ref{BH050}) and~(\ref{BH060}) into the nonlinear
Schr{\"o}dinger equation in imaginary space and time. Therefore, Eqs.~(\ref{BH050}) and~(\ref{BH060})
belong to a family of completely integrable models. The only problem (but potentially a big one) is to
adapt the ISM to a finite system with periodic boundaries and to accommodate the problem-specific boundary
conditions~(\ref{BH040}) and~(\ref{eq:final-constr}). The exact solution would provide
a full analytic control of the subleading corrections to the action of branch 2, which are presently half-empiric.

It would be very interesting to explore the possibility of extending to the spatially averaged KPZ
interface height some of the recent ``stochastic integrability" approaches, which led, for selected initial
conditions, to exact representations for the complete statistics of the \emph{one-point} interface
height~\cite{SS,CDR,Dotsenko,ACQ,CLD11,CLD12,IS12,IS13,Borodinetal}.

Finally, we can try to put our results into a broader perspective of DPTs
in large deviations of macroscopic systems far from equilibrium. The second-order DPT in the present periodic KPZ system
occurs when the unstructured optimal height profile $h(x,t) = \bar{H} t$, uniformly translating in the vertical direction, gives way to spatially non-uniform height profiles via a supecritical bifurcation. This transition can be compared with second-order DPTs in large deviations of the current \cite{Bertinietal,BodineauD,HG,ZarfatyM} and of the system activity \cite{Lecomte} in some diffusive lattice gas models with periodic boundaries. In each of these systems  a spatially uniform state gives way to a non-uniform state via a supercritical bifurcation. The structured states, however, have the form of simple travelling waves, whereas in the present system they are more complicated.

A second-order DPT is also observed in large deviations of the \emph{one-point} height of a KPZ interface in infinite system
with Brownian initial conditions \cite{Janas2016,KLD2017,SKM2018,KLD2022}. The underlying bifurcation is also supercritical, but the transition mechanism is quite different. Indeed, the optimal height profiles $h(x,t)$ are non-uniform and essentially non-stationary both below and above the transition, while the transition is accompanied by a mirror symmetry breaking.

Similarly, the first-order DPT, that we reported here,  differs from its counterpart  in large deviations of the one-point height of a KPZ interface \emph{at a shifted point} in the KPZ equation with Brownian initial conditions  \cite{SKM2018}.
These differences between mechanisms of different DPTs of the same order call for a better understanding of DPTs in general.


\section*{Acknowledgments}
The authors thank Eldad Bettelheim and Naftali R.\ Smith for useful discussions.
This research was supported by the program
``Advanced Research Using High Intensity Laser-Produced Photons and Particles"
(ADONIS) (CZ.02.1.01/0.0/0.0/16019/0000789) of the European Regional Development Fund (ERDF) (PS),
and by the Israel Science Foundation (Grant No. 1499/20) (BM).


\appendix
\section{Numerical methods}
Our numerical procedure of finding solutions $h$ and $p$ of the
OFM problem~(\ref{BH050})-(\ref{eq:final-constr})
can be summarized as follows:
To compute numerical solutions to the boundary-value problem
for $h$ and $p$ for given $\ell$ and $\bar{H}$, we use a
refined version of the popular Chernykh--Stepanov
back-and-forth iteration algorithm~\cite{CS} as described in detail
in Ref.~\cite{SGMG}, using the language of PDE-constrained optimization.
The idea is to interpret the back-and-forth
iterations -- fixing $\Lambda$ and solving Eq.~(\ref{BH050}) forward in time
with fixed $p$, and Eq.~(\ref{BH060}) backward in time with fixed $h$ until
convergence -- as adjoint~\cite{Plessix} gradient evaluations $\delta S /
\delta p$ of the action
functional with fixed $\Lambda$,
\begin{equation}
S[p] = \frac{1}{2} \int_0^1 dt \int_0^\ell
d x \, p^2(x,t) - \Lambda
\int_0^\ell h[p](x,1) dx\,,
\label{eq:action-lbda}
\end{equation}
with the height profile $h = h[p]$ determined for a
given $p$ through Eq.~(\ref{BH050}).
This interpretation allows us to use automatic update step-size
control (here: Armijo line search~\cite{Armijo}) and
preconditioning for faster convergence (here: L-BFGS method~\cite{LN}).
Conceptually, one fixes $\Lambda$ in this formulation and obtains
the corresponding average height value $\bar{H}$ \textit{a posteriori}.

For large $\ell$ we find multiple solutions for the
same $\bar{H}$, and the action $S(\bar{H},\ell)$ of the optimal solution as a
function of $\bar{H}$
becomes nonconvex for some $\bar{H}$. Nonconvexity of the rate
function $S(\bar{H})$ is an issue because
minimizing the functional~(\ref{eq:action-lbda}) effectively computes the
Legendre--Fenchel transform of the rate function at $\Lambda$,
which may diverge in this case. Therefore, we add a
penalty term to the action, leading to the so-called
augmented Lagrangian formulation~\cite{Hestenes}
\begin{equation}
S[p] = \frac{1}{2} \int_0^1 dt \int_0^\ell
d x \, p^2(x,t) - \Lambda \left(
\int_0^\ell h[p](x,1) dx  - \ell \bar{H}\right)
+ \frac{\mu}{2} \left(\int_0^\ell h[p](x,1)
dx  - \ell \bar{H}\right)^2\,,
\label{eq:action-lbda-mu}
\end{equation}
and solve multiple minimization problems for increasing penalty
parameters $\mu$.
In this formulation, one can directly prescribe $\bar{H}$ at the
cost of solving multiple optimization problems, and it is usable
regardless of convexity of the rate function, or in other words regardless of
bijectivity of the map between $\bar{H}$ and $\Lambda$.

The formulation~(\ref{eq:action-lbda}) is more convenient to
trace solution branches: one initializes the optimization on an
already found solution on a given branch and slightly changes
$\Lambda$. In order to trace branches close to the transition
region for large $\ell$ in
the nonconvex case, we temporarily reparameterize the observable
as described in Ref.~\cite{AG} with reparameterizations
$g(z) = \ln \ln z$ or $g(z) = 1 - \exp \{-(z - 3.5) \}$.

Within this general framework, we use a
pseudo-spectral code with spatial resolution $n_x$
to solve Eqs.~(\ref{BH050})
and~(\ref{BH060}), with an exact integration of the diffusion
terms through an integrating factor in Fourier space. An explicit
second-order Runge--Kutta integrator with $n_t$ equidistant steps
is used in time. The gradient of the action functional is
evaluated exactly on a discrete level (``discretize,
then optimize''). A Python source code which illustrates the optimization
methods in a simple toy problem
can be found in a public GitHub repository~\cite{STGS-gh}, and it is
explained in Ref.~\cite{STGS}.

An important property of Eqs.~(\ref{BH050}) and~(\ref{BH060}) is their exact integrability \cite{Janas2016,KLD2021,KLD2022}. It is convenient to exploit it for monitoring the accuracy of our numerical discretization scheme. To this end we
followed in time the first five conserved quantities of the continuous system~\cite{KLD2021}:
\begin{align}
c_1 &= \int p \; dx\\
c_2 &= \int p \partial_x h \; dx\\
c_3 &= \int p \left[\partial_{xx} h + \frac{1}{2} \left(\partial_x h \right)^2 + \frac{1}{2}p \right] \; dx\\
c_4 &= \int p \left[\partial_{xxx} h + \frac{3}{2} \partial_x h \partial_{xx} h + \frac{1}{4} \left(\partial_x h \right)^3 + \frac{1}{2} \partial_x p + \frac{3}{4} p \partial_x h \right] dx\\
c_5 &= \int p \bigg[\partial_{xxxx} h + \frac{3}{2} (\partial_{xx} h)^2 + \frac{1}{8} (\partial_x h)^4 + 2 \partial_x h \partial_{xxx} h + \frac{3}{2} (\partial_x h)^2 \partial_{xx} h + \frac{1}{8} \partial_x h \left[2 \partial_x p + 3 p \partial_x h \right] \nonumber\\
&\quad + \frac{1}{4} \left[2 \partial_{xx} p + 3 \partial_x p \partial_x h + 3 p \partial_{xx} h \right] + \frac{1}{8} p (\partial_x h)^2 + \frac{1}{2} p \left[\partial_{xx} h + \frac{1}{2}(\partial_x h)^2 + \frac{1}{2} p \right]\bigg] dx
\end{align}
for the optimal paths that we calculated numerically. The code conserves $c_1$ up to machine precision, but the higher $c_i$ are only approximately conserved. For instance for the optimal path shown in the top row of Fig.~\ref{fig:branch-2-3-solutions}, for the nonzero quantities $c_3$ and $c_5$, we obtain conservation within $0.3
\%$ and $0.6\%$, respectively, for our numerical solution. 
For other optimal paths at different $\bar{H}$ and $\ell$, we observe a similar, sufficiently high,  accuracy of the conservation of $c_3$ and $c_5$ in our numerical results. (The quantities $c_2$ and $c_4$ should theoretically be equal to $0$ for all times due to the initial condition $h(x, t= 0)= 0$ and periodic boundary conditions in space. They are less convenient, therefore, for the analysis of relative accuracy of numerical results.)

Due to the fixed and small time interval $0\leq t\leq 1$, intrinsic to the problem of short-time statistics of the KPZ interface, there are no adverse consequences of the violation of exact conservation of the quantities $c_i$. This violation would be a more serious problem if we would attempt to advance the solution until long times, $t\gg 1$. While for this reason we have not derived an integrable discretization in space and time, such discretizations do exist. They are described in detail \textit{e.g.}\ in the monograph~\cite{Surisbook}.


\end{document}